\documentclass{aa}  
\usepackage{graphicx}

\usepackage{amsmath}

\usepackage{hyperref}

\usepackage[colorinlistoftodos]{todonotes}

\usepackage{txfonts}

\begin{document} 
   \title{Axisymmetric Schwarzschild models of an isothermal axisymmetric mock dwarf spheroidal galaxy.}
   %\titlerunning{Axisymmetric Schwarzschild models of a mock dwarf spheroidal galaxy.} % 2-column format
   \titlerunning{Axisymmetric models of a mock dwarf spheroidal galaxy.} % referee format (shorter running title needed)
   % isothermal vs isotropic
   \subtitle{}

   \author{Jorrit H.J. Hagen
          %\inst{1}
          \and
          Amina Helmi %\inst{1}
          \and
          Maarten A. Breddels %\inst{1}
            %\fnmsep\thanks{Just to show the usage of the elements in the author field}
          }

   \institute{Kapteyn Astronomical Institute, University of Groningen, P.O. Box 800, 9700 AV Groningen, The Netherlands \\
\email{hagen@astro.rug.nl}
        %\and
        %     University of Alexandria, Department of Geography, ...\\
        %     \email{c.ptolemy@hipparch.uheaven.space}
        %     \thanks{The university of heaven temporarily does not
        %             accept e-mails}
             }

   \date{Received 27 Jun 2019 / Accepted DD Mon YYYY}

% \abstract{}{}{}{}{} 
% 5 {} token are mandatory
 
  \abstract
  % context heading (optional)
  % {} leave it empty if necessary  
   {}
  % aims heading (mandatory)
   {The goal of this work is to test the ability of Schwarzschild's
     orbit superposition method in measuring the mass content, scale
     radius and shape of a flattened dwarf spheroidal
     galaxy.  Until now, most dynamical model efforts have assumed
     that dwarf spheroidal galaxies and their host halos are
     spherical.}
     %, even though their observed light distribution is clearly not spherical.}
  % methods heading (mandatory)
   {We use an Evans model (1993) to construct an isothermal mock
     galaxy whose properties somewhat resemble those of the Sculptor dwarf
     spheroidal galaxy. This mock galaxy contains flattened luminous
     and dark matter components, resulting in a logarithmic profile
     for the global potential. We have tested how well our Schwarzschild
     method could constrain the characteristic parameters of the
     system for different sample sizes, and also if the functional form of
     the potential was unknown.}
  % results heading (mandatory)
   {When assuming the true functional form of the potential, the
     Schwarzschild modelling technique is able to provide an accurate
     and precise measurement of the characteristic mass parameter of
     the system and reproduces well the light distribution and the
     stellar kinematics of our mock galaxy. When assuming a different
     functional form for the potential, such as a flattened NFW
     profile, we also constrain the mass and scale radius to their
     expected values. However in both cases, we find that the
     flattening parameter remains largely unconstrained. This is
     likely because the information content of the velocity dispersion
     on the geometric shape of the potential is too small, since
     $\sigma$ is constant across our mock dSph.}
  % conclusions heading (optional), leave it empty if necessary 
   {Our results using Schwarzschild's method indicate that the mass
     enclosed can be derived reliably, even if the flattening parameter is unknown, 
     and already for samples containing $2000$
     line-of-sight radial velocities, such as those currently
     available. Further applications of the method to more general
     distribution functions of flattened systems are needed to
     establish how well the flattening of dSph dark halos can be
     determined.}


   \maketitle
%
%________________________________________________________________

\section{Introduction}

In the current cosmological $\Lambda$CDM model most of the mass is
believed to be in the form of (cold) dark matter. While successful on
large scales, on the scales of dwarf galaxies, the model suffers a
number of challenges, including the missing satellites problem
\citep{Klypinetal1999, Mooreetal1999}, the cusp-core conundrum
\citep{Hui2001}, and the too big to fail problem
\citep{Boylan-Kolchin2011}, although all may be solved one way or
another by considering the effects of baryonic physics \citep[e.g.][]{Zolotovetal2012,Brooksetal2013, Wetzeletal2016, Kimetal2018}.  The dwarf spheroidal satellite galaxies (dSph’s or dSph galaxies) of our Milky
Way can provide particularly strong constraints on the nature of dark
matter, since their high mass-to-light ratios suggest that they are
fully dark matter dominated
\citep{Strigarietal2008,Walkeretal2007,Wolfetal2010}. 

Various methods have been used to develop dynamical models of dSph
galaxies using line-of-sight velocity measurements for large samples
of individual stars in these systems \citep[e.g.][]{Battagliaetal2006, Battagliaetal2008, Battagliaetal2008b, Battagliaetal2011, Walkeretal2009, Walkeretal2015_Draco}. Modelling
via the Jeans Equations, distribution functions, and orbit
superposition methods like Schwarzschild modelling are amongst those
most often used \citep{Battagliaetal2013}. All these methods have in common that they assume that
the systems are in dynamical equilibrium. 

The Jeans Equations are derived by taking moments of the Collisionless
Boltzmann Equation, which itself describes the conservation of
probability in phase-space \citep{BinneyTremaine2008}. Not every
solution of the Jeans equations has an associated distribution
function that is physical (i.e. positive) everywhere. Furthermore
finding a solution requires additional assumptions, for example on the
functional form of the density profile and on the velocity anisotropy
\citep[because this is generally not known, although see the work
by][who determined directly the anisotropy of a sample of stars in the
Sculptor dSph using proper motions derived from {\it Gaia} and
HST]{Massarietal2018}. Because Jeans modelling is very flexible and fast
it has become the most widely used tool to model dSph galaxies,
particularly in the spherical limit. It has, for example, allowed for a
robust (independent of the velocity anisotropy) measurement of the
mass enclosed within approximately the half light radii of the dSph
galaxies \citep{Walkeretal2009b, Wolfetal2010}, and the determination
that the masses of the classical dSph's are in the range
$\sim(10^8-10^9) M_{\odot}$ \citep[e.g.][]{Walkeretal2007}. On the
other hand, it has not been possible to rule out cusped or cored
profiles on the basis of these types of models
\citep[e.g.][]{Evansetal2009,Strigarietal2017}.

The \citet{Schwarzschild1979} modelling technique relies on the idea
that a system can be seen as a superposition of stellar orbits. In
Schwarzschild modelling one only needs to assume a specific
gravitational potential form. The method does require a significant
amount of computing power and therefore a smaller set of gravitational
potentials can be explored in comparison to Jeans modelling.
\citet{BreddelsHelmi2013} have applied this method to 4 dwarf
spheroidal galaxies and by modelling both the second and fourth
line-of-sight velocity moments and assuming spherical symmetry they find that, 
independently of the particular form assumed for the potential, it is
possible to constrain not only the mass at around the half-light
radius (more precisely at $r_{-3}$ where the logarithmic slope of the
luminous density is $-3$) but also the logarithmic slope of the
dark matter density.

Most work thus far has assumed that dwarf spheroidal galaxies and
their host halos are spherical, despite the fact that their light
distribution is typically not round \citep{Irwin&Hatzidimitriou1995,
  McConnachie2012}.  Furthermore, dark matter halos are predicted to
be triaxial \citep{Jing&Suto2002} when no baryonic effects are taken
into account, although subhalos in cold dark matter simulations that
could host dSph's are only mildly triaxial, and almost axisymmetric
\citep{Veraetal2014}. This implies that it is important to establish
how many and which of the previously mentioned results still stand
when taking into account deviations from spherical symmetry.

\citet{Kowalczyketal2017_recoveringmassandanisotropy,
  Kowalczyketal2018_theeffectofnonsphericity} have in fact studied the
ability of recovering the mass profile and anisotropy of the remnants
of the mergers of dwarf disky galaxies (one postulated channel for the
formation of dSph) when using spherical Schwarzschild models. These
authors have shown that for spherical remnants the method can break
the mass-anisotropy degeneracy, whereas for non-spherical (prolate)
remnants the anisotropy will always be underestimated, although the
total mass profile will be recovered well for data along the minor
axis (although not if the data are along the major axis). 

On the other hand, \citet{Hayashi&Chiba2012, Hayashi&Chiba2015} used
axisymmetric Jeans modelling to infer the axis ratio of the dark
matter density distribution ($Q$) in several dSph's assuming a
constant velocity anisotropy $\beta_z$. They report rather low axis
ratios ($Q=[0.3-0.5]$) compared to the observed projected flattening
in the light ($q^\prime_* \sim 0.7$). These low values are somewhat
counterintuitive, though the results may be affected by degeneracies
between $Q$, the velocity anisotropy profile, the viewing angle of the
dSph, and the inner slope of the dark matter density profile. In
\citet{Hayashietal2016}, a very similar technique was applied to
unbinned data, and for e.g. Scl dSph, the authors found that the
flattening parameter is largely unconstrained. 

In this work we explore the performance of the Schwarzschild modelling
technique in the axisymmetric regime, to free ourselves from the
assumptions inherent to Jeans models. We test the method on a mock
Sculptor-like dSph and consider axisymmetric mass distributions for
both the light and the dark matter component and establish how well
the characteristic parameters of the potential can be recovered, for
different sample
sizes. 

The paper is organised as follows. In Sect. \ref{sec:amockgalaxy}, we set up a mock galaxy and simulate a realistic dataset. In Sect. \ref{sec:method} we describe the Schwarzschild method and its implementation in this work. Then, in Sect. \ref{subsec:recoveringthemockgalaxyparameters}, we apply the Schwarzschild method and show that we can recover the characteristic mass parameter of the mock galaxy potential, irrespective of the potential flattening assumed. In Sect. \ref{subsec:nfwmodels} we model our mock galaxy with an axisymmetric NFW potential form and show that, even in this case, the Schwarzschild method is able to constrain the mass and scale radius to the expected values for datasets containing a realistic number of stars. We present our conclusions in Sect. \ref{sec:discussionandconclusion} where we also discuss our findings.

%__________________________________________________________________
%__________________________________________________________________
%__________________________________________________________________

\section{The mock galaxy}
\label{sec:amockgalaxy}

%__________________________________________________________________
%__________________________________________________________________
%__________________________________________________________________

\subsection{Potential, luminous density and characteristic parameters}
\label{subsec:potential}

We have built a mock galaxy inspired in the Sculptor dSph. We have
assumed a flattened stellar density profile ($q_{\ast}=0.8$), no net
rotation and a line-of-sight velocity dispersion of $\sim 10$~km/s
\citep{Mateo1998,Battagliaetal2008, Walkeretal2009b}. For simplicity,
we have set up the mock galaxy following \citet{Evans1993}, who uses
an elementary distribution function to describe a composite
axisymmetric system. This distribution function is ergodic, i.e. it
leads to a velocity ellipsoid that is isotropic and has a constant
amplitude and thus is not generic\footnote{nor is this distribution
  function ideal as we shall see later in the paper, because it
  provides very little information on the symmetries of the system.}.

The gravitational potential of the composite system as a whole follows the form
\begin{equation}
\label{eq:logarithmicpotential}
\Phi_\text{E}(R,z) = \frac{1}{2} v^2_0 \, \text{ln} \left( R^2_c + R^2 + \frac{z^2}{q^2} \right) + \Phi_0 \, ,
\end{equation}
where ($R$, $\phi$, $z$) denote the cylindrical coordinates. Here
${v_0}^2$ relates to the mass of the system and $R_c$ is the core
radius. The parameter $q$ is the axial ratio, and has to satisfy
$1/\sqrt{2} = 0.707 \leq q \leq 1.08$ where the lower limit is set by
the condition that the spatial density is positive everywhere
\citep{BinneyTremaine2008} and the upper limit yields a composite
distribution function of the form used by \citet{Evans1993} that is
positive everywhere. The zero point of the potential is set by
$\Phi_0$.

The density profile of the stellar component is described by
\begin{equation}
\label{eq:rholum}
\rho_\text{lum}(R,z) = \frac{\rho_0 R^p_c}{ \left( R^2_c + R^2 + z^2 / q_{\ast}^2 \right)^{p/2} } \, ,
\end{equation}
where $\rho_0$ is the central density, $p$ denotes a slope parameter, and $q_{\ast}$ is the flattening of the stellar density. The associated stellar distribution function is given by
\begin{equation}
\label{eq:flum}
f_{\text{lum}}(E) \propto \text{exp}[-pE/v_0^2] = \text{exp}[-p\Phi_\text{E}/v_0^2] \,\, \text{exp}[-pv^2/2v_0^2] \, ,
\end{equation}
where $E$ is the sum of the gravitational potential and kinetic energies of a star.

In the Evans model $q_{\ast} = q$ and therefore the density flattening
of the luminous component is the same as the potential (not the
density) flattening of the composite system. The surface brightness
profile of the mock galaxy can be found by integrating the luminous
density along the line-of-sight.

The line-of-sight velocity profile is exactly Gaussian with a velocity dispersion that is isotropic and constant everywhere:
\begin{equation}
\label{eq:sigmaevans}
\sigma_{\text{E}} = \frac{v_0}{\sqrt{p}} \, ,
\end{equation}
and independent of the inclination, scale radius and flattening. 

We choose here $v_0=20$~km/s, $R_c=1$~kpc, $q=0.8$, and $p=3.5$ for
our mock galaxy. These values result in a velocity dispersion of
roughly $10.7$~km/s.  For these values of $p$ and $q$, the central
total density should be at least $1.13$ times the central stellar
density to yield positive phase-space densities for
both the stellar and dark components everywhere.

\begin{figure}[ht!] \centering 
 \includegraphics[width=0.5\textwidth]{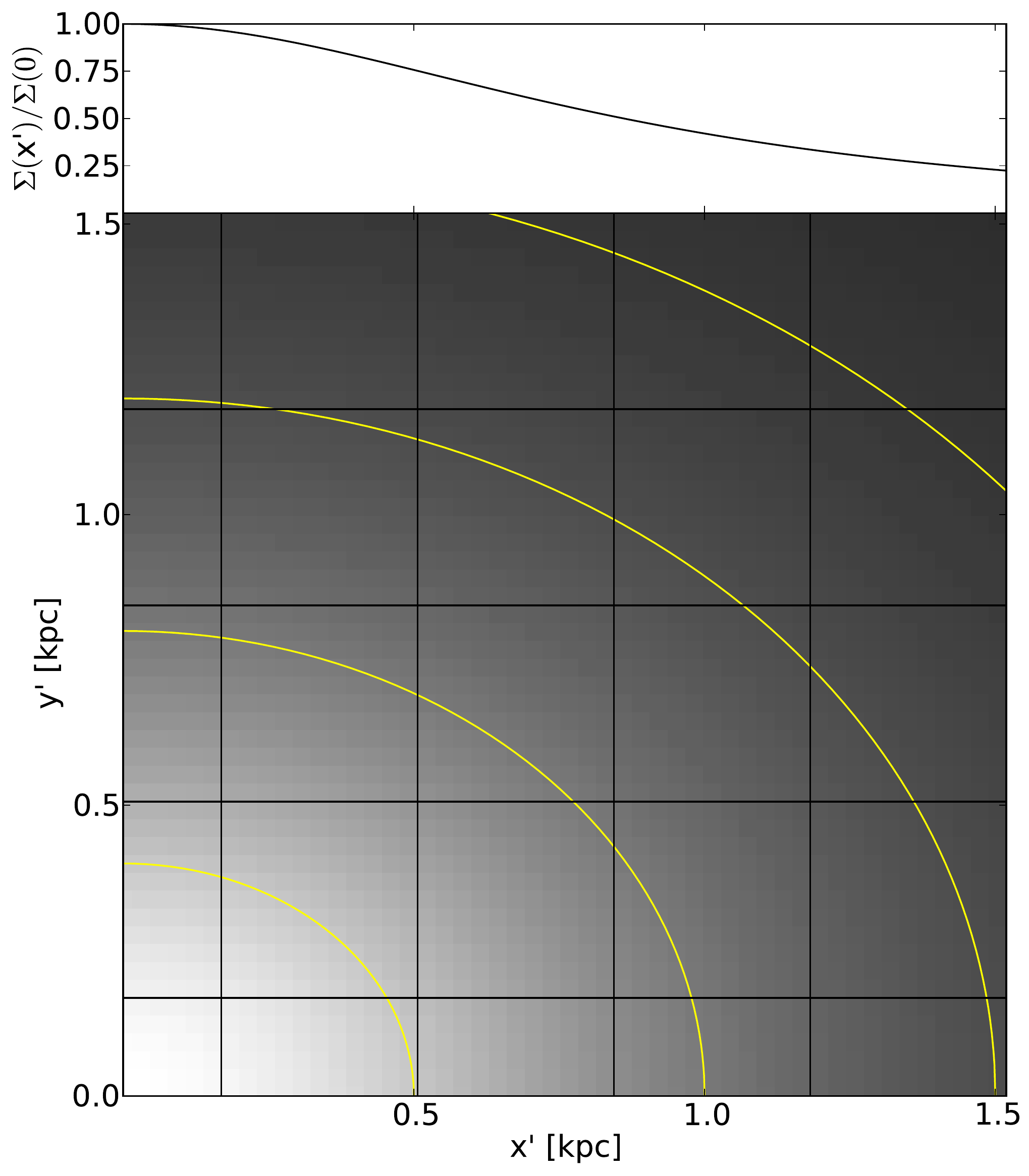} 
 \caption{The surface brightness profile of our mock galaxy in an
   edge-on view.  The black horizontal and vertical lines show the
   boundaries of the kinematic-bins.  We only show the positive
   quadrant of our FOV ($x'>0$~kpc, $y'>0$~kpc). The yellow contours
   correspond to the isophotes of the system ($q_{\ast}=q=0.8$).  In
   the top panel we have plotted the surface brightness normalised to
   its central value as function of $x'$, i.e. along the (projected)
   major axis of the galaxy.}
 \label{fig:mockgalaxy} \end{figure}
 
In Fig. \ref{fig:mockgalaxy} we show the 2D surface brightness profile
of the mock galaxy for an edge-on view. Since the galaxy is
axisymmetric, we only show the positive quadrant. Contours of constant
surface brightness follow ellipses with axial ratios $q'_{\ast}$,
which because of the edge-on view are identical to the intrinsic
density flattening (i.e. $q'_{\ast}=q_{\ast}=0.8$). The 1D surface
brightness profile along the major axis is plotted in the top panel of
this figure. The surface
brightness decreases a factor two with respect to its central value at
a projected ellipsoidal radius of $0.86$~kpc, however, the projected
half light radius is much larger (\mbox{3.87 kpc}).

%__________________________________________________________________
%__________________________________________________________________
%__________________________________________________________________

\subsection{Observing the mock galaxy}
\label{subsec:observing}

We generate the mock galaxy by drawing positions following the
luminous density distribution (see Eq.  \ref{eq:rholum}) and
velocities from the Gaussian distribution function (see
Eq. \ref{eq:flum}). We thus assume that the dataset of the stars with
line-of-sight velocities follows the same distribution as the
light. We place the mock galaxy at a distance of $80$~kpc, and
``observe'' it with a square field of view (FOV), centred on the mock
galaxy, with a size of $7832^{\arcsec} \times 7832^{\arcsec}$ (which
then corresponds to roughly $3 \times 3$~kpc). Throughout this work we
assume an edge-on view.

The typical line-of-sight velocity measurements of individual stars
have errors of order $\mathrm{d}v = 2$ km/s
\citep{Battagliaetal2008b,Walkeretal2009}.  Therefore, to simulate a
realistic dataset we convolve the line-of-sight velocities with a
Gaussian distribution having a standard deviation of $2$~km/s.  We
compute velocity moments by combining the velocities of all available
stars in a certain spatial bin on the sky (in what follows a {\it
  kinematic-bin}) in our FOV. The velocity moments are estimated by
correcting for the measurement errors (see Appendix
\ref{App:AppendixA}), similarly to \citet{Breddelsetal2013}. We assume
that the surface brightness profile can be measured without error in
much smaller spatial bins on the sky (which we refer to as {\it
  light-bins}). To produce a reasonable galaxy, we also assume that
the three-dimensional light distribution is known to much larger
radii, but for many fewer bins (more details can be found in
Sect.~\ref{subsec:generatingorbitlibraries}).

%__________________________________________________________________
%__________________________________________________________________
%__________________________________________________________________

\section{The Schwarzschild orbit superposition method}
\label{sec:method}

In Schwarzschild modelling, orbits are used as building blocks of a
dynamical system. Given a potential $\Phi$, a complete set of orbits
are integrated numerically and for each orbit the predicted
observables are stored in a so-called orbit library. Varying the
parameters of the potential (or varying the potential form as a
whole), will result in different libraries.  The library which
provides a combination of weighted orbits that matches the
observations (light profile + kinematics) best, will be said to yield
the best-fit parameters of the potential. The orbital weights
themselves provide the corresponding distribution function. Since the
orbital weights are positive by construction, the distribution
function will be non-negative everywhere.

%__________________________________________________________________
%__________________________________________________________________
%__________________________________________________________________

\subsection{Generating orbit libraries}
\label{subsec:generatingorbitlibraries}

In this paper we use a slightly modified version of the Schwarzschild
code from \citet{vandenBoschetal2008}, who modelled the elliptical
galaxy NGC4365. In what follows, we shortly describe how we generated
the orbit libraries, how the orbital integration has been done and how
the libraries are stored. For more information we refer the reader to
\citet{vandenBoschetal2008}\footnote{Note that the Schwarzschild code
  by \citet{vandenBoschetal2008} was developed to model triaxial
  systems, and therefore also generates initial conditions for box
  orbits, which have zero time-averaged angular momentum and which can
  cross the centre \citep{Schwarzschild1979,Schwarzschild1993}. In an
  axisymmetric potential $L_z$ is conserved and such box orbits will
  therefore never attain velocities in the azimuthal direction. As
  this could cause non-axisymmetries in our model we do not 
  specifically generate box orbits.}.

Given an energy $E_i$, initial positions ${x_0}$ and ${z_0}$ are
sampled on a open polar grid, which is defined by $N_{I_2}$ polar
angles and $N_{I_3}$ radii in between a thin orbit and the
equipotential. The polar angles are sampled linearly, but to obtain a
better sampling of orbits near the major axis of the system, 50\% of
the polar angles are sampled from the $z$-axis towards $10^\circ$
above the midplane, and the remaining 50\% from $10^\circ$ down to the
$z=0$ plane.  The initial $y$-coordinates and initial velocities in
the $x$- and $z$-directions are set to zero.  The initial velocities
in the $y$-direction, ${v_{y,0}}$, are determined by
\mbox{$E_i - \Phi({x_0},{0},{z_0}) = 0.5 {v_{y,0}}^2$}.  This is done
for all $N_\text{ener}$ energies, which are defined by
\mbox{$E_i=\Phi(x=x_i ,y=0, z=0)$}. The locations $x_i$ that fix the
energy grid are logarithmically sampled between $25$~pc and $50$~kpc
from the centre.  This `orbit library’ thus consists of
\mbox{$N_\text{orb} = N_\text{ener} \times N_{I_2} \times N_{I_3}$}
orbits ($z$-tubes in our axisymmetric potential).

To account for slowly precessing orbits in the library we also compute
$17$ copies of each orbit, where each copied orbit is subsequently
rotated by $10^\circ$ in the $xy$-plane. These $18$ copies are summed
into a single orbit and replace the non-rotated orbit, such that each
orbit now follows the axisymmetric requirements. Besides ensuring
axisymmetric behaviour of our models, adding rotations also increases
the sampling of an orbit. Note as well that each orbit has a
counter-rotating sibling, obtained by appropriately changing the sign
of the velocity vector.

We further improve the accuracy of the model by `dithering': every
orbit is split into $N^3_\text{dither}$ suborbits by replacing each of
its three nonzero initial coordinates by $N_\text{dither}$ slightly
different coordinates. In fact, the initial conditions of all
suborbits are found by increasing $N_\text{ener}$, $N_{I_2}$, and
$N_{I_3}$ by a factor of $N_\text{dither}$.  The observables of each
set of adjacent $N^3_\text{dither}$ suborbits are combined and stored
as being the observables of the (bundled) orbit. Choosing an odd
number for $N_\text{dither}$ ensures that the original orbit is the
central suborbit of the bundle. In all our Schwarzschild models we use
$N_\text{dither}=5$. Every main orbit is thus made from a bundle of
$5^3=125$ neighbouring suborbits.

We use a Runge Kutta integrator to compute the stellar trajectories
over roughly $200$ orbital time scales. We require that the energy of
each suborbit is always conserved better than $1\%$ by increasing the
accuracy of the integrator if necessary.  For each orbit the kinematic
information is stored in a velocity grid, which consists of a
line-of-sight velocity axis ($N_\text{v}$ velocity bins) and an axis
associated to the location on the sky ($N_\text{kin}$ kinematic-bins).
On equally spaced time intervals, a count is added to the element of
the grid associated to the velocity and location at the given time.
The sky projected path of the orbit is determined in a similar way and
stored in the surface brightness grid containing $N_\text{2D light}$
light-bins. In an additional 3D grid containing $N_\text{3D light}=800$ bins
($40$ radial, $5$ azimuthal and $4$ polar bins in the positive octant)
the 3-dimensional path of an orbit is stored. This 3D grid reaches
radii well beyond the FOV (in contrast to the velocity and surface
brightness grid), and is used to control the system at such radii.

In this work we set $N_\text{2D light}$ equal to $99 \times 99 = 9801$
and $N_\text{kin}$ to $9 \times 9 = 81$, unless stated otherwise.  The
velocity axis of the velocity grid contains $N_v = 41$ bins and has a
total velocity width of $80$~km/s, such that we cover velocities up to
$\pm 4 \sigma_{\text{E}}$. The central velocity bin is centred on
$0$~km/s. To be able to track how long an orbit spends in a given
kinematic-bin, counts will also be added to the first or last velocity
bin if velocities are beyond the limits of the velocity
grid\footnote{When taking too few (i.e. too wide) velocity bins for
  the velocity grid, the velocity moments might not be recovered
  correctly. We have also checked that if we bin the true Gaussian
  line-of-sight velocity profile of our mock galaxy as described
  above, thus discarding the contribution of velocities that are
  outside the range of the grid, the velocity moments are recovered
  well, i.e. the first and third moments are not affected, while the
  second and fourth velocity moments might result in relative errors
  of order 0.1\% and 2\% given the choices made for binning the
  velocity data.}.

%__________________________________________________________________
%__________________________________________________________________
%__________________________________________________________________

\subsection{Fitting orbital weights}
\label{fitting}\

Once the orbit libraries are in place, we find the orbital weights
such that the total luminous mass, the surface brightness profile and
the kinematics within the FOV, and the 3D light profile of the system
are reproduced.

The 2D light profile is fitted using the surface brightness grid, where we define:
\begin{equation}
m^{\text{light}}_j = \sum\limits_{i=1}^{N_\text{orb}} w_i \, m^{\text{light}}_{ij} \, ,
\end{equation}
where we sum over all orbits $i$. Here, $m^{\text{light}}_{ij}$ is the fraction of time orbit $i$ spent in light-bin $j$ and $m^{\text{light}}_j$ is the fractional surface brightness in light-bin $j$. The orbital weights are denoted by $w_i$ and add to unity by construction. The 3D light profile is fitted similarly using the 3D grid.

At the same time as we fit the light, we also fit the kinematics. In
every kinematic-bin $k$ we compute the first 4 mass-weighted velocity
moments $\langle v^n_k \rangle$ by defining:
\begin{equation}
m^{\text{kin}}_k \langle v^n_k \rangle = \sum\limits_{i=1}^{N_\text{orb}} w_i \, m^{\text{kin}}_{ik} \langle v^n_{ik} \rangle \, ,
\end{equation}
where again we sum over all orbits $i$. This time $m^{\text{kin}}_{ik}$ is the fraction of time orbit $i$ spent in kinematic-bin $k$ and $m^{\text{kin}}_k$ is the fractional surface brightness in kinematic-bin $k$. The $n^{\text{th}}$ moment of orbit $i$ in kinematic-bin $k$ is given by $\langle v^n_{ik} \rangle$:\\
\begin{equation}
\langle v^n_{ik} \rangle  = \frac{\displaystyle \sum\limits_{l=2}^{N_v-1} h_{ikl} \, v_{\text{cen},l}^n \, \triangle v} {\displaystyle \sum\limits_{l=2}^{N_v-1} h_{ikl} \, \triangle v} \, ,
\end{equation}
where, $\triangle v$ is the size of the velocity bin and $h_{ikl}$ is
the fraction of time that orbit $i$ spent in kinematic-bin $k$ and
velocity bin $l$. Velocity bin $l$ has velocity range
[$v_{\text{cen},l} - \frac{1}{2} \triangle v$,
$v_{\text{cen},l} + \frac{1}{2}\triangle v$], where $v_{\text{cen},l}$
denotes its central velocity. We sum over the $N_v$ velocity bins,
although we discard the contributions of the first and last velocity
bin. This is done since we did not set a stringent outer velocity
boundary in these velocity bins: as described before, counts will be
added here even if a star has a velocity outside the range of the grid
and therefore the typical velocities of these bins are not known. Note
that $m^{\text{kin}}_{ik} = \sum\limits_{l=1}^{N_v} h_{ikl}$ with this
choice.

Now that we have defined the relation between the observables and the quantities in our model, we can describe how we fit the orbital weights.
The fit is based on minimising $\chi^2_{\text{tot}}$:
\begin{equation}
\label{eq:chi2total}
\chi^2_{\text{tot}} = \sum\limits^{N_\text{obs}}_{u=1} \left[ \frac{\text{Model}[u] - \text{Data}[u]}{\text{Error}[u]} \right]^2,
\end{equation}
where $u$ runs over all $N_\text{obs}$ observables. The number of
observables is given by:
\begin{equation}
\label{eq:Ntotal}
N_\text{obs} = 1 + N_\text{2D light} + N_\text{3D light} + 4N_\text{kin},
\end{equation}
which includes the contribution of the total light of the system, the fractional light for each 2D and 3D light-bin, and the four velocity moments for each kinematic-bin, respectively. Since using higher order moments might reduce the degeneracy between the velocity anisotropy and the mass profile \citep[e.g.][]{Merrifield&Kent1990,Richardson&Fairbairn2013} we choose to use four velocity moments. We do not use higher moments since these are observationally harder to constrain.

We use a non-negative least square solver to ensure that all orbital weights are positive. The light is weighted by assigning an error of 2\% to each of the 2D and 3D light-bins.

We note that we can investigate the individual contribution to the total $\chi^2_{\text{tot}}$ by decomposing it, e.g:
\begin{equation}
\label{eq:chi2tot}
\chi^2_{\text{tot}} = \chi^2_{\text{total light}} + \chi^2_{\text{2D light}} + \chi^2_{\text{3D light}} + \chi^2_{\text{kin}}.
\end{equation}
We stress that $\chi^2_\text{tot}$ is being minimised. We do not
minimise the terms on the right-hand side individually. The term
associated to the total light of the system turns out to be
negligible, since it is always recovered very well. The
same holds for $\chi^2_{\text{3D light}}$. These terms are only added to Eq. \ref{eq:chi2total}
to ensure that the model returns a realistic galaxy (in the sense that the luminous component might resemble a galaxy).
Most of the constraining
power thus comes from the surface brightness profile and the
kinematics.

%__________________________________________________________________
%__________________________________________________________________
%__________________________________________________________________

\section{Results}
\label{results}
In this section we show that the Schwarzschild method can recover some
of the characteristic parameters of the mock Sculptor-like dwarf
spheroidal galaxy. We first show in Sect.
\ref{subsec:recoveringthemockgalaxyparameters} that if the true
potential functional form of the system is known, we can constrain the
characteristic mass parameter of the mock galaxy.  In reality however,
the true potential functional form is not known. Therefore, in Sect.
\ref{subsec:nfwmodels} we demonstrate how well we can constrain the
characteristic parameters when assuming an axisymmetric form of a
Navarro-Frenk-White \citep[NFW,][]{NFW1996} potential.

%__________________________________________________________________
%__________________________________________________________________
%__________________________________________________________________

\subsection{Two parameter Evans models: recovering the mock galaxy parameters}
\label{subsec:recoveringthemockgalaxyparameters}

\begin{figure}[t!] \centering 
 \includegraphics[width=0.5\textwidth]{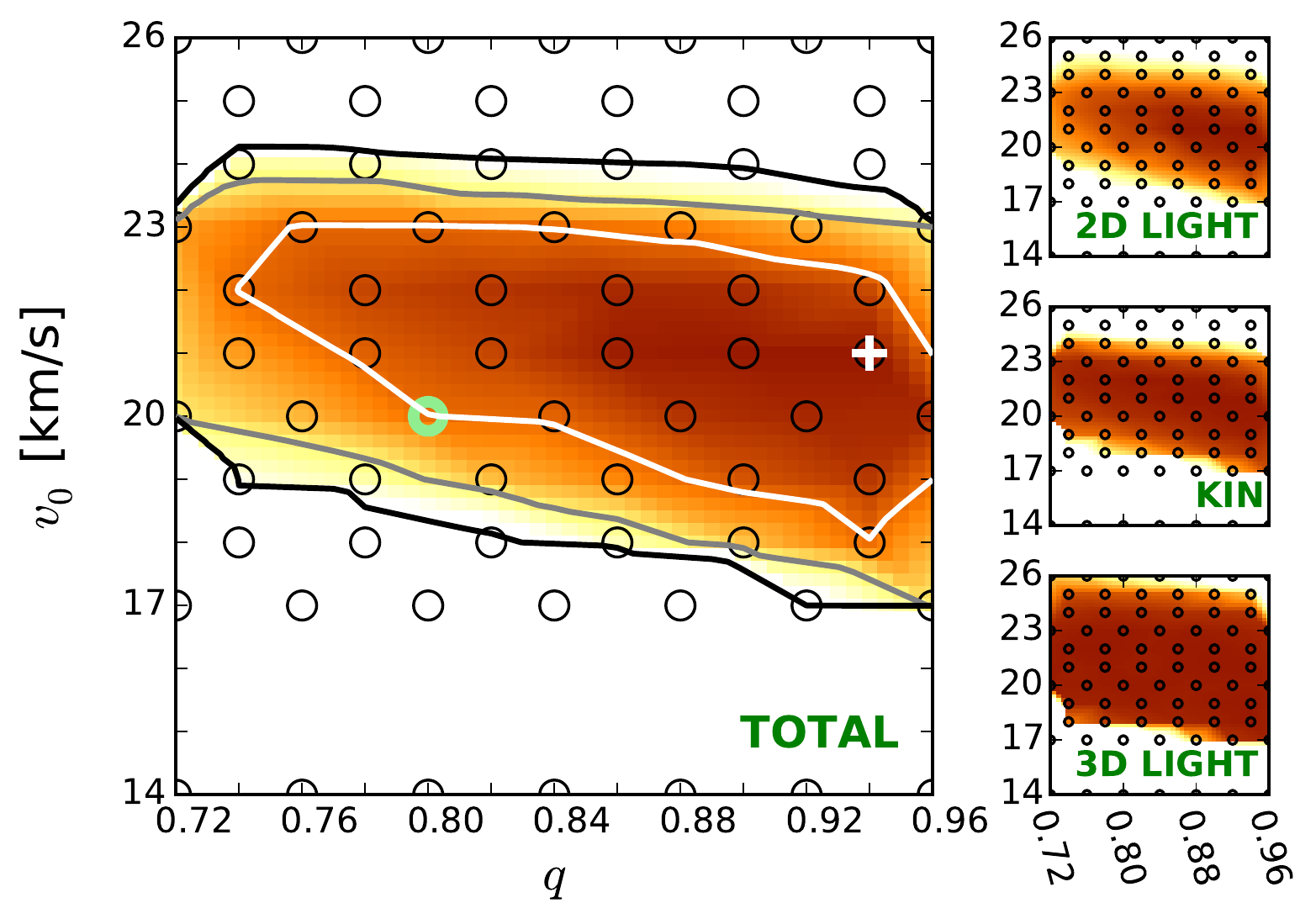} 
 \caption{
 $\Delta \chi^2$-distribution 
 %Probability distribution 
 of the characteristic parameters $q$ and $v_0$ of the Evans models obtained after applying the Schwarzschild method. In this case our mock data consist of $10^5$ stars inside the FOV \mbox{($3 \times 3$~kpc)}. We use 9x9 kinematic-bins and assume the potential functional form and inclination are known. The black circles show the locations where the Schwarzschild models were evaluated. The green circle indicates the input parameters of the mock system. The best-fit model is indicated by the white cross and recovers the mock galaxy mass parameter.
 In white, grey and black we show the $\Delta \chi^2=[2.3, 6.18, 11.8]$-contours respectively. 
 The coloured landscape shows interpolated $\Delta \chi^2$-values, and goes up to a maximum of $\Delta \chi^2=10$. On the right we show the $\Delta \chi^2$-landscapes when only considering $\chi^2_{\text{2D light}}$ (top), $\chi^2_{\text{kin}}$ (middle), or $\chi^2_{\text{3D light}}$ (bottom).
 } 
 \label{fig:evans_results1_probcontours} \end{figure}
 
Here we assume the true form of the potential is known, i.e. we use it
to build the orbit libraries for the Schwarzschild models. Our aim is
to establish whether we can recover the correct values of the
characteristic input parameters with this method. To this end we make a grid of models
in which we vary the values of the characteristic parameters $q$ and
$v_0$ (see Eq.  \ref{eq:logarithmicpotential}). We thus fix the core
radius to \mbox{$R_c = 1$ kpc}, i.e. to its true value. We sample $q$
from $0.72$ to $0.96$, and $v_0$ from $11$~km/s to $29$~km/s, with
higher sampling (decided iteratively) having steps in $q$ of 0.02 and in
$v_0$ of $1$~km/s. We name the models by the values of their
parameters: qXXvYY in which XX = $100q$ and YY = $v_0$ in km/s.  For
the orbit sampling, we set $N_\text{ener}=32$, $N_{I_2}=32$ and
$N_{I_3}=16$ such that a total of
$32 \times 32 \times 16 \times 5^3 = 2048000$ suborbits are integrated
(see Sect. \ref{subsec:generatingorbitlibraries}) and
$2 \times 32 \times 32 \times 16 = 32768$ orbital weights are
determined (see Sect. \ref{fitting}).

%__________________________________________________________________
%__________________________________________________________________
%__________________________________________________________________
\begin{figure}[t!] \centering
\includegraphics[width=0.45\textwidth]{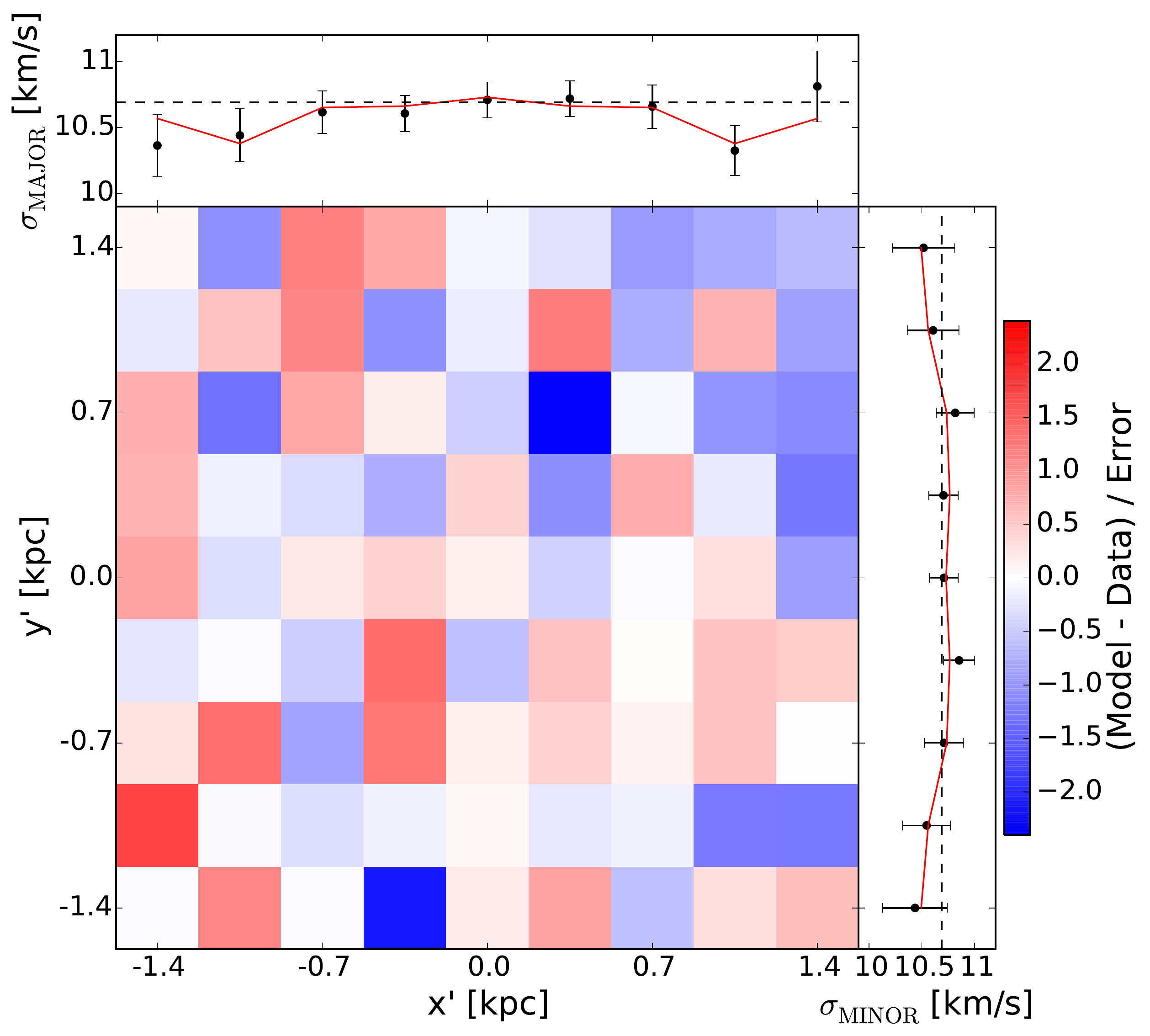}
\caption{The difference of the best fit and the observed velocity dispersion in terms of the observed error, for all 9x9 kinematic-bins. The figure is obtained after fitting the q94v21 library to our mock data consisting of $10^5$ stars in our FOV, assuming an edge-on view. The top and the right panels show the fit (red full line) obtained along the major and minor axis respectively. The data points with 68\% error bars are shown in black. Black dashed lines indicate the true velocity dispersions from theory (Eq. \ref{eq:sigmaevans}).}
\label{fig:testevans_2apertures_100000k_moments_fov_theon_results1_q80v20_chisigma}  \end{figure}

\subsubsection{Results for a large sample}
\label{subsubsec:resultsforalargesample}

We start with an idealised case in which the data consist of $10^5$ stars. For 9x9 kinematic-bins on the sky, the typical error of the velocity dispersion in a kinematic-bin is $\sim 0.25$~km/s. 

The large panel of Fig. \ref{fig:evans_results1_probcontours} shows the results obtained by fitting the Schwarzschild models to the data.
The small black circles show the grid of tested values for $q$ and $v_0$, the green circle the true input values, and the white cross indicates the values of the parameters corresponding to the maximum likelihood estimator (MLE). For the best-fit model q94v21 we find $\chi^2_\text{tot} = 207.7$. The contribution of the kinematics (see Eq. \ref{eq:chi2total}) to this value is $205.6$. Using 81 kinematic-bins to fit 4 velocity moments, this corresponds to $0.64$ per kinematic constraint. 

We have computed
$\Delta \chi^2(q, v_0) = \chi^2_\text{tot}(q, v_0) -
\textrm{min}[\chi_\text{tot}^2]$ for each of these models and define
68\%, 95\% and 99.7\% -confidence intervals (white, grey and black
contours, respectively) at $\Delta \chi^2=[2.3, 6.18, 11.8]$
\citep{Pressetal1992}\footnote{We used the
  scipy.interpolate.LinearNDInterpolator to interpolate the
  $\Delta \chi^2$.}. The coloured background shows the
$\Delta \chi^2$-landscape and is truncated at $\Delta \chi^2 = 10$.
The smaller panels on the right show the $\Delta \chi^2$-landscapes
when only considering $\chi^2_{\text{2D light}}$ (top),
$\chi^2_{\text{kin}}$ (middle), or $\chi^2_{\text{3D light}}$
(bottom). The $\Delta \chi^2$-landscape based on $\chi^2_{\text{tot}}$
is thus slightly dominated by the differences in
$\chi^2_{\text{2D light}}$, although the kinematics provide similar
constraints. 

To estimate the error on the mass parameter we first marginalize over
the flattening parameter by selecting for each $v_0$ the minimum
$\Delta \chi^2$ along $q$. We define the 68\% error at
those values where $\Delta \chi^2 = 1.0$ \citep{Pressetal1992}. For
this experiment we find \mbox{$v_0 = 21^{+1.33}_{-2.11}$ km/s}. We
therefore conclude that we can recover the input mass parameter of our
mock galaxy well, but as Figure~\ref{fig:evans_results1_probcontours}
shows we do not constrain well the flattening $q$.

In
Fig. \ref{fig:testevans_2apertures_100000k_moments_fov_theon_results1_q80v20_chisigma}
we show how well the velocity dispersion is fitted in the best-fit
q94v21 model. For each kinematic-bin, we show how much the model
deviates from the data expressed in units of the error on the data. On
top we show the fit along the major axis while the subpanel on the
right shows the fit along the minor axis. These figures show that the
fit is very good (and in fact, it is almost indistinguishable from
the fit obtained for what would be the input parameters model, i.e. q80v20).

%__________________________________________________________________
%__________________________________________________________________
%__________________________________________________________________

\subsubsection{Downsampling and folding data}
\label{subsubsec:downsamplingandfoldingdata}

We now consider the more realistic case of a sample of $10^4$
stars. To reduce the observed uncertainties on the kinematics we
decided to fold the kinematic data (but not the light).  Since the
system is axisymmetric, we fold our data into the kinematic-bins
located in the first quadrant.  We can simply move each star towards
its corresponding kinematic-bin without changing its velocity, because
our system has an identical Gaussian line-of-sight profile everywhere
(see Sect. \ref{subsec:potential}). In general, however, one should
change the velocities following the assumed symmetry.

Since we fold the data from 9x9 bins of our FOV into the first
quadrant, we effectively have $10^4$ stars located in the resulting
5x5 kinematic-bins. A typical kinematic-bin now contains $400$ stars
on average, and the typical error on the velocity dispersion is
$\sim 0.45$ km/s.

\begin{figure}[t!] \centering 
	\includegraphics[width=0.5\textwidth]{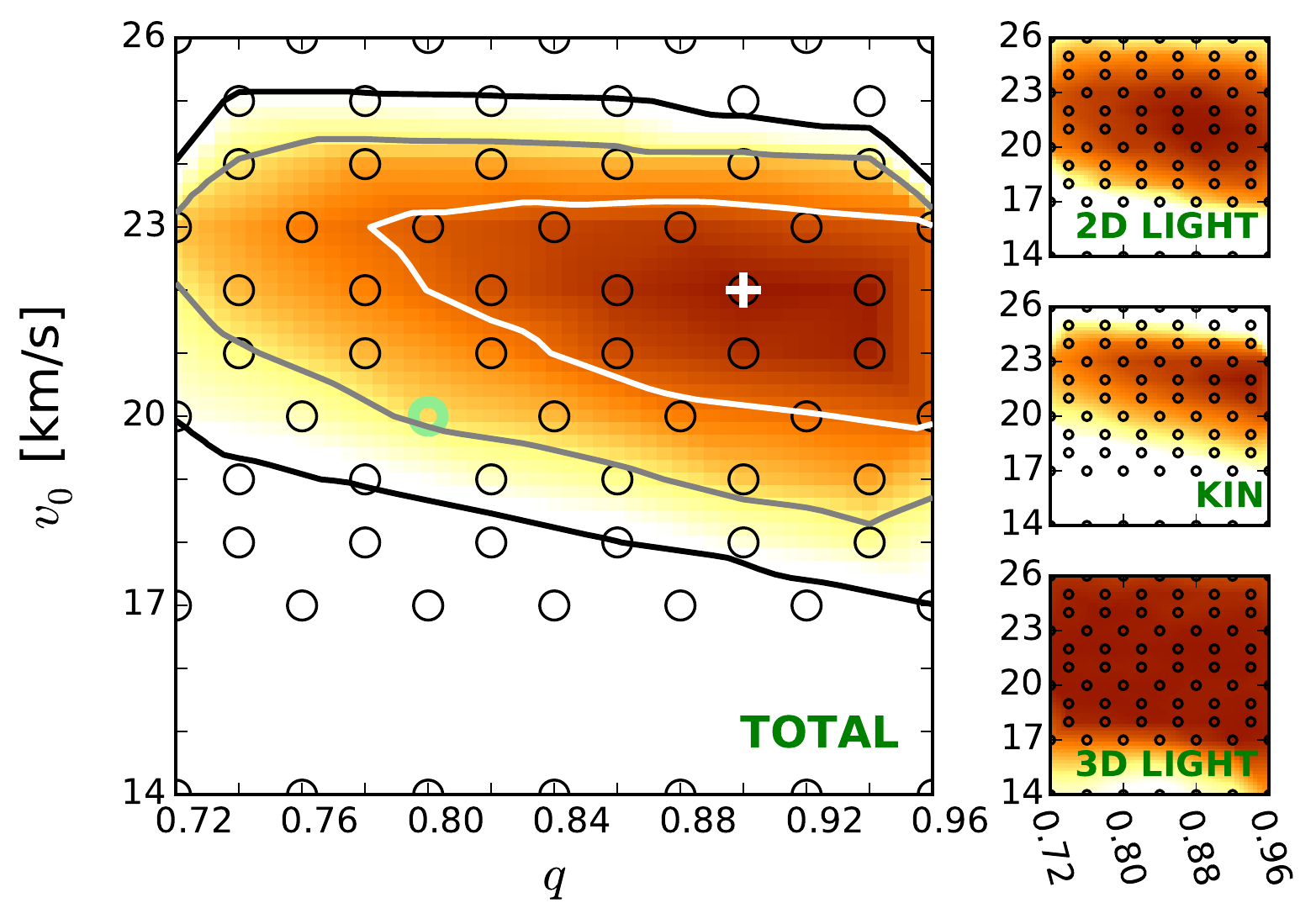} \caption{Similar to Fig. \ref{fig:evans_results1_probcontours}, but now using $10^4$ stars and using the approach of folding the data from 9x9 into 5x5 kinematic-bins. The parameter inferences are similar, though slightly larger masses are preferred.} 
	\label{fig:evans_results2_folded9x9_probcontours} \end{figure}

We fit the folded data with the Schwarzschild orbit superposition
method and find the MLE for model q90v22 (see
Fig. \ref{fig:evans_results2_folded9x9_probcontours}). As in the case
of $10^5$ stars, and thus as expected, the flattening parameter
remains fairly unconstrained. We find a slightly larger mass parameter
\mbox{$v_0 = 22^{+1.02}_{-1.44}$ km/s}, but \mbox{$v_0 = 20$ km/s} is
still within the 95\%-confidence region.

For the best-fit model q90v22 we find $\chi^2_\text{tot} = 16.5$. The contribution of the kinematics (see Eq. \ref{eq:chi2total}) to this value is $13.2$. Both values are much lower than in the case of $10^5$ stars,  
and this can be explained by the decrease in the number of kinematic constraints and the fact that the data have now been folded.

It is encouraging that a more realistic number of stars still gives
such tight constraints. Comparing the $10^4$ stars folded case to the
case of $10^5$ stars, the 2D 68\%-probability contours are shifted
towards just slightly larger masses. Note that the uncertainty on the
mass parameter did not increase.

\begin{figure}[t!] \centering 
	\includegraphics[width=0.5\textwidth]{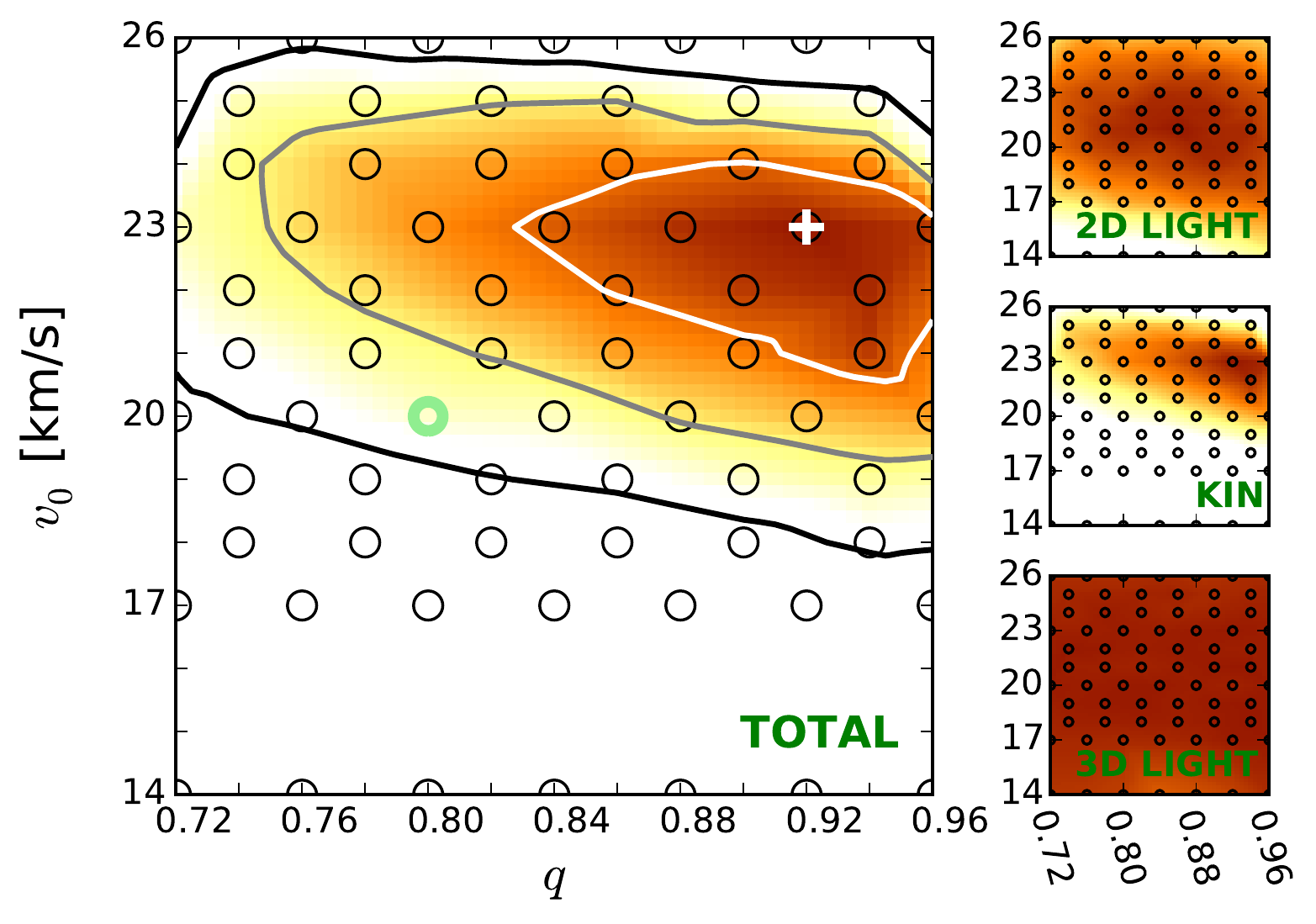} \caption{Similar to Fig. \ref{fig:evans_results2_folded9x9_probcontours}, now using $2000$ stars. The parameter inferences are similar, though slightly larger masses are inferred ($\Delta v_0 \sim 1$~km/s).} 
	\label{fig:evans_results2000_folded9x9_probcontours} \end{figure}

To further test how the results depend on the number of stars observed, we decreased even further the number of stars to a sample of $2000$ stars. This is the typical size of currently available datasets used to put constraints on the mass of dSph galaxies \citep[e.g.][]{Walkeretal2009b,BreddelsHelmi2013,Hayashi&Chiba2015}. We again fold the data from 9x9 into 5x5 kinematic-bins. The resulting typical error on the velocity dispersion in a kinematic-bin is then $\sim 0.9$~km/s.
In this case we find a best-fit model q92v23 ($\chi^2_\text{tot} = 38.9$, $\chi^2_\text{kin} = 32.6$). The $\Delta \chi^2$-distribution is shown in Fig. \ref{fig:evans_results2000_folded9x9_probcontours}. The best models are again reproduced by the most round models, although statistically the flattening parameter remains unconstrained. The region spanned by the contour drawn at $\Delta \chi^2=11.8$ is of similar size, but is shifted towards slightly higher masses ($\Delta v_0 \sim 1$~km/s) in comparison to the case of $10^4$~stars. The true q80v20 model is nevertheless still within the inferred 99.7\%-confidence interval. 

The weak trend found for smaller samples to prefer slightly higher
values of $v_0$ may be due to the fact that, for small radii (compared
to $R_c$), the potential (see Eq.~\ref{eq:logarithmicpotential}) is proportional to
$v_0^2 [\ln R^2_c + (R/R_c)^2] + (v_0/q)^2 (z/R_c)^2$. Therefore,
there is a weak degeneracy in the term $v_0/q$, that may manifest
itself more when the sampling is sparse, and thus lead to a small
shift in preferred values of $v_0$ for larger $q$. 

      \hfill \break From the tests performed in this Section we
      conclude that, with a kinematic sampling that follows the light,
      we can not aim to constrain the flattening of an {\it
        isothermal} dSph galaxy\footnote{Slightly better results can
        be obtained by sampling uniformly with distance, see
        Appendix~\ref{app:light}.}, even if the true functional form
      of the potential is known. This is likely because the
      information content in a velocity dispersion regarding the
      geometric shape of the potential is too small (since $\sigma$ is
      constant across the whole system). We can however still reliably
      constrain the mass parameter of such a system, i.e. even though
      the true flattening remains unknown. This can already be done
      for a realistic number of stars.

%__________________________________________________________________
%__________________________________________________________________
%__________________________________________________________________

\subsection{Axisymmetric NFW models}
\label{subsec:nfwmodels}

We have shown that the Schwarzschild method can constrain correctly the mass parameter when the true functional form of the potential is known. Now, we will tackle the problem more realistically by allowing a different functional form for the potential. We consider an axisymmetric NFW-profile, and follow the parametrization of \citet{Vogelsbergeretal2008}:
\begin{equation}
\label{Vogelsbergerpotential}
\Phi_\text{V}(\tilde{r}) = -4\pi G \rho_0 R^3_s \left[ \frac{\ln(1+\tilde{r}/R_s)}{\tilde{r}} \right] \, , 
\end{equation}
where $R_s$ is the scale radius and $\rho_0$ a characteristic density parameter. In comparison to the spherical NFW-profile, the radius $r=\sqrt{R^2 + z^2}$ is replaced by a newly defined radius:
\begin{equation}
\label{eq:rtilde}
\tilde{r}= \frac{(r_a+r)r_E}{r_a+r_E} \, , 
\end{equation}
where, for the axisymmetric case, $r_E= \sqrt{\left(\frac{R}{a}\right)^2 + \left(\frac{z}{c}\right)^2}$ is the ellipsoidal radius with $a$ and $c$ specifying the relative lengths of the major and minor axes, and where $r_a$ is a transition radius. In addition, we require that $2a^2+c^2=3$, such that when $a=c=1$, this results in the spherical NFW profile. For $r>>r_a$, $\tilde{r} \rightarrow r$, whereas for $r<<r_a$, $\tilde{r} \rightarrow r_E$. Therefore, the gravitational potential is axisymmetric in the central regions and becomes spherical in the outer regions. We set the transition radius to $r_a = 10$ kpc. In all our Vogelsberger models we keep the transition radius $r_a$ fixed.

To additionally guarantee that the total mass density is positive up to at least the orbits possessing the highest energies in our library ($\sim 50$ kpc), the flattening parameter must satisfy $c/a \gtrsim 0.7$ for a case with \mbox{$R_s = 1$ kpc}. For smaller scale radii, larger lower limit values of $c/a$ are needed to satisfy the positive density criterion. 

For convenience, we define a characteristic mass parameter, $M_\text{1kpc}$ expressed in units of $M_{\odot}$, which corresponds to the total enclosed mass within 1 kpc from the centre for a spherical NFW profile with scale radius $R_s$, i.e. 
\begin{equation}
\label{NFWmass}
M_\text{NFW}(r = 1 {\rm kpc} \, \vert \, R_s) = 4 \pi \rho_0 {R_s}^3 \left[ \ln \left(\frac{R_s + r}{R_s} \right) - \left( \frac{r}{R_s + r} \right) \right]_{1 {\rm kpc}} \, .
\end{equation}
From this equation we determine the value of $\rho_0$, and it is this
value of $\rho_0$ that we use for the axisymmetric Volgelsberger potential
in Eq.~\ref{Vogelsbergerpotential}.

%__________________________________________________________________
%__________________________________________________________________
%__________________________________________________________________

\subsubsection{The `true' (equivalent) Vogelsberger system}
\label{subsubsec:truevogelsbergersystem}
Before we can test the Schwarzschild orbit superposition method while assuming Vogelsberger mass models, we need to know when a result can be considered satisfactory. Since we could not constrain the flattening for the case when the true potential form is known we will not aim to constrain the flattening for the Vogelsberger models. Nevertheless, the expected best-fit scale radius $R_s$ and mass $M_\text{1kpc}$ of our system will depend on the $c/a$-value assumed. 
In this section we therefore establish what are good parameters for the mass $M_\text{1kpc}$, scale radius $R_s$, and flattening $c/a$, such that the properties of the Evans mock galaxy are reproduced the best. 

\begin{figure*}[ht!] \centering
    \includegraphics[width=0.497\textwidth]{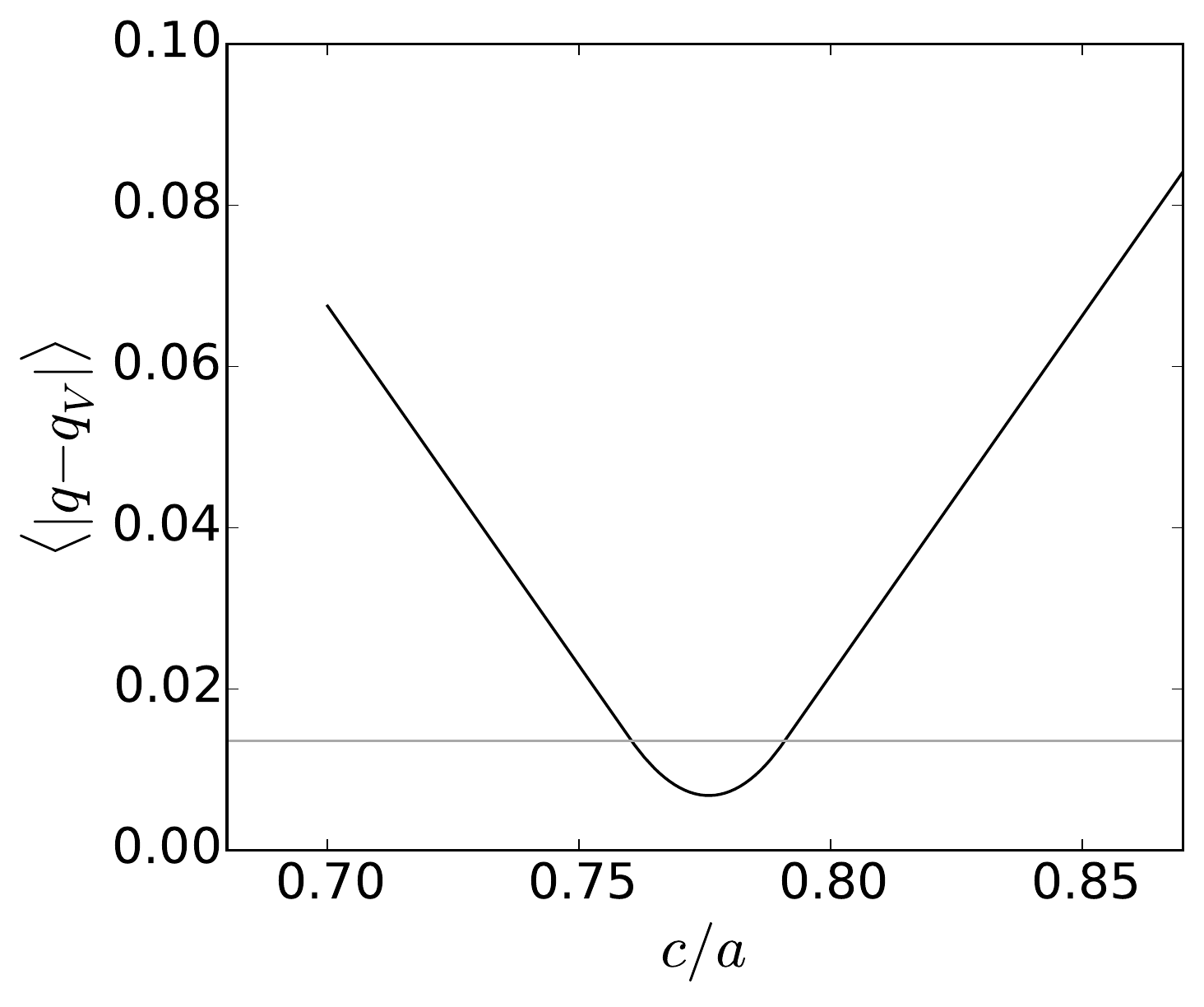}%
	\includegraphics[width=0.5\textwidth]{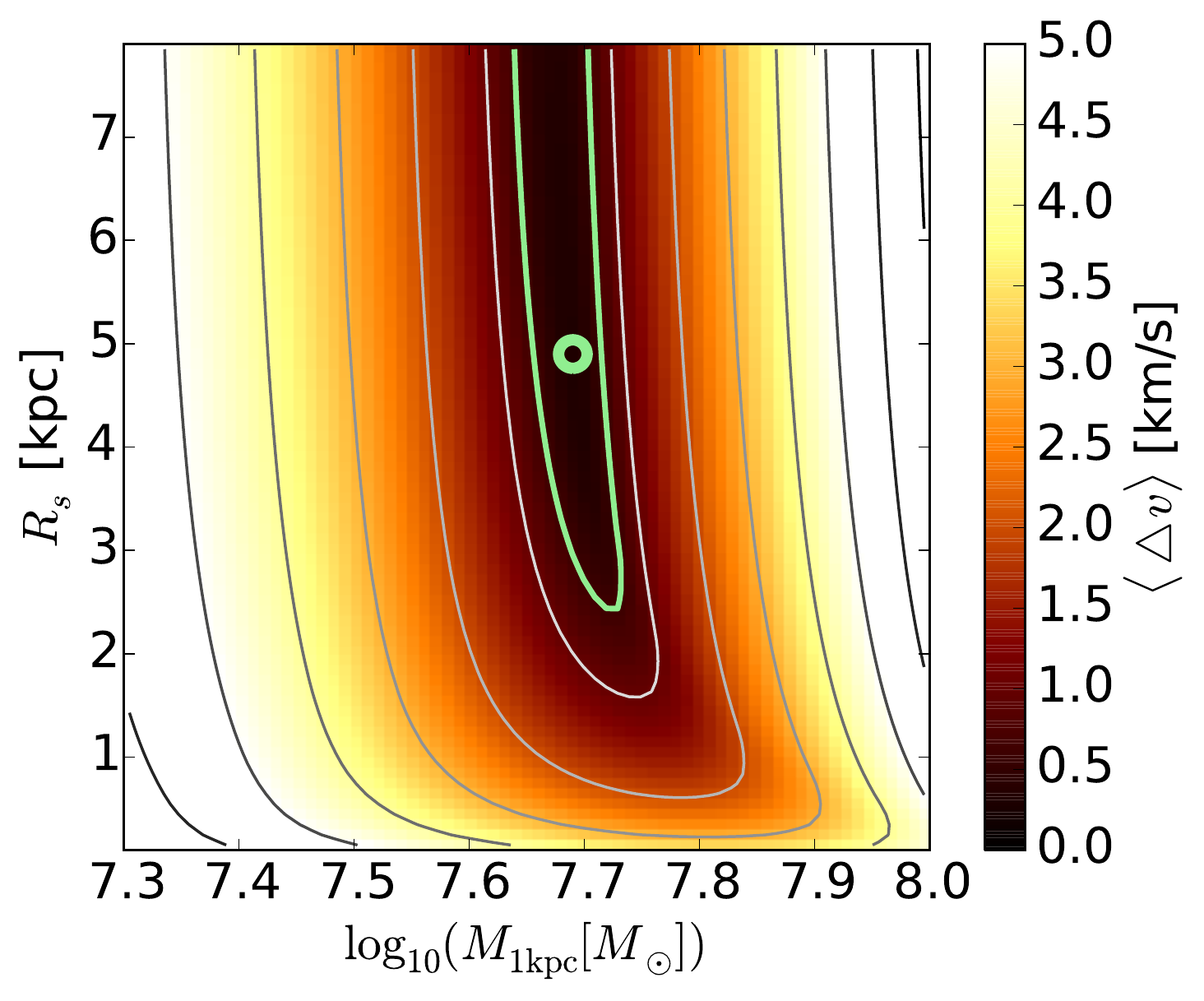}%
	\caption{Estimating the `true' parameters of the Vogelsberger system by comparing the differences in potential flattening on the left and the differences in the gradients of the potentials on the right. The comparisons are based on the distance interval from $0.5$ up to $2.0$~kpc (with steps of $0.05$~kpc) from the centre of the galaxy. Left: The mean absolute difference of the Vogelsberger potential flattening and the true potential flattening of the mock galaxy as a function of the flattening parameter $c/a$ of the Vogelsberger potential (black line). 
    The grey horizontal line marks the positions where this difference has doubled, with respect to the minimum $0.007$ at $c/a \simeq 0.776$.
	Right: We minimise the mean of the absolute differences in the gradients of the potential along the major and minor axis (compared to the mock galaxy) by varying the Vogelsberger model parameters $M_\text{1kpc}$ and $R_s$. The figure is obtained after setting the flattening parameter to $c/a = 0.776$. The colour bar is truncated at $5.0$~km/s. The green circle indicates the location at $\log_{10}(M_\text{1kpc}[M_{\odot}]) \simeq 7.69$ and $R_s = 4.9$~kpc where the differences are minimum ($\langle \bigtriangleup v \rangle_{\rm min} = 0.31$~km/s). Grey lines indicate the contours of constant mean absolute differences and are spaced by $1$~km/s. As a proxy for the error on the Vogelsberger parameters, a green contour is drawn where the differences are doubled with respect to the minimum difference.} \label{fig:findingvogelsbergertruth} \end{figure*}%

% Figures Vogelsberger Truth
\begin{figure*}[ht!] \centering
	\includegraphics[width=0.55\textwidth]{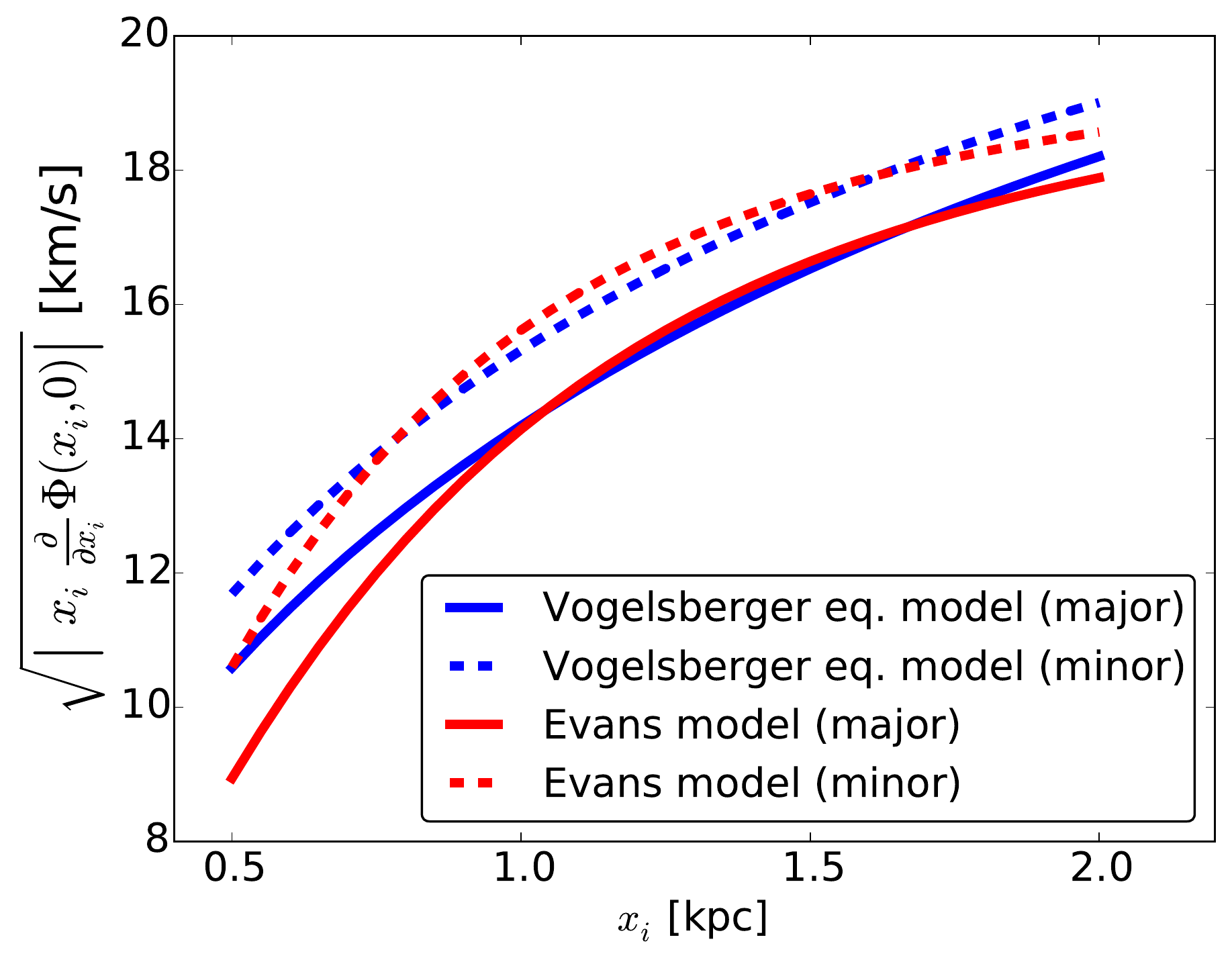}%
    \includegraphics[width=0.44\textwidth]{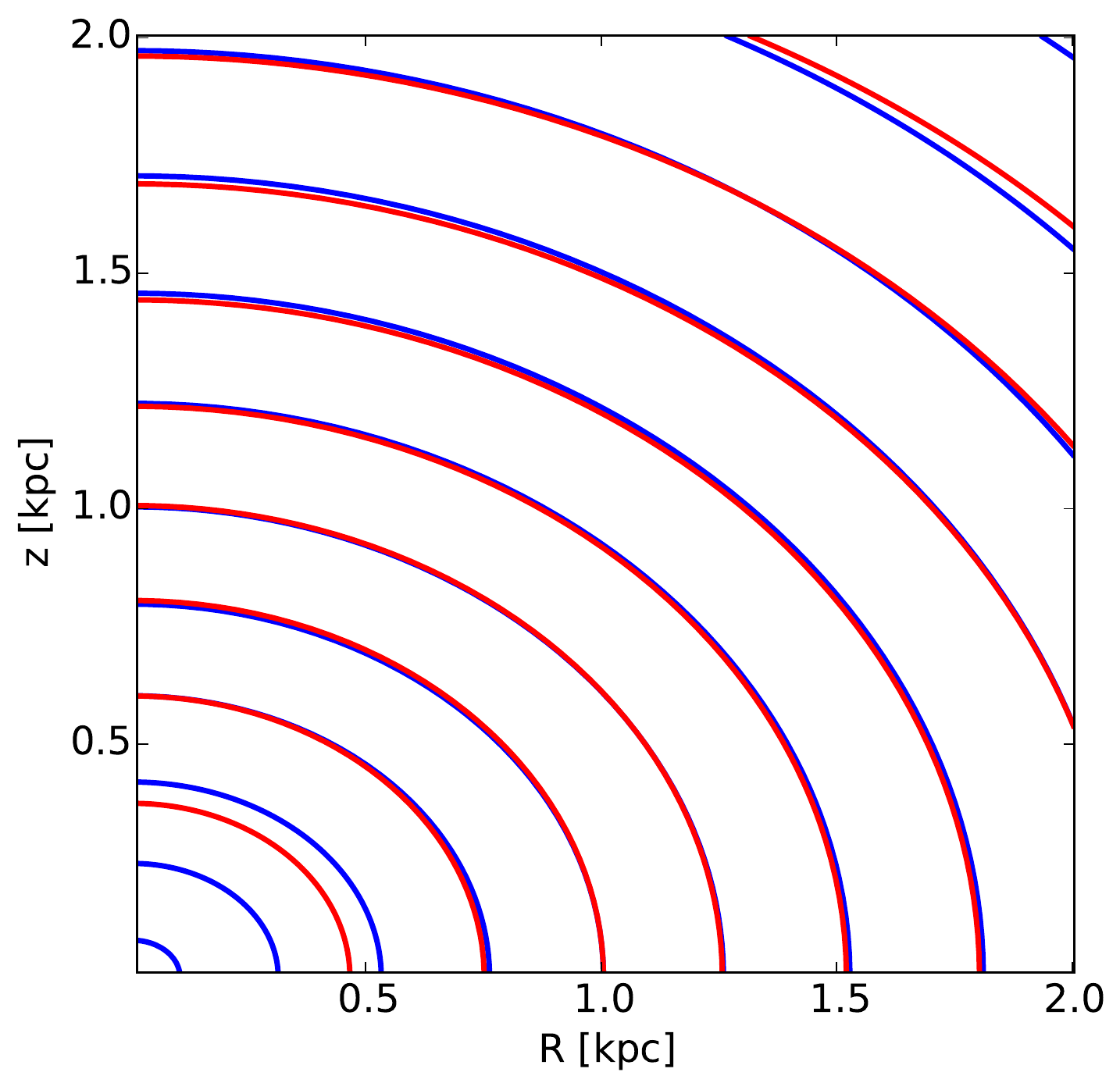}%
    \caption{Left: The major (full lines) and minor (dashed lines)
      axis gradients of the potential as function of $R$ and $z$
      respectively for the true Evans model (red) and for the
      ``equivalent'' Vogelsberger potential with $c/a \simeq 0.776$,
      $R_s \simeq 4.9$~kpc and
      $\log_{10}(M_\text{1kpc}[M_{\odot}]) \simeq 7.69$ (blue).
      Right: Comparison of the isopotential contours for the true
      Evans (red) and the equivalent Vogelsberger models (blue). For
      the purpose of this figure the zero-point of the potential is
      chosen here such that $\Phi_\text{E}=\Phi_\text{V}$ at
      $(R,z)=(1,0)$~kpc. For each potential, contours are drawn at the
      positions where $\Phi$ has changed in steps of
      $50$~km$^2$/s$^2$.  In the region from $0.7$ up to $2$~kpc the
      equivalent Vogelsberger model follows well the true Evans
      potential. For more inner radii the (cusped) Vogelsberger models
      can not reproduce the (less steep) cored behaviour of the Evans potential.
    } \label{fig:comparingvogelsbergertruth} \end{figure*}%
	
  Because most stars of our mock galaxy will have projected radii in
  between $0.5$ and $2.0$~kpc from the centre, we require that the
  flattening of the Vogelsberger potential should be comparable to
  that of the mock galaxy over this region.  At a given position we
  define the Vogelsberger potential flattening $q_\text{V}$ as the
  axis ratio of the equipotential contour that goes through that
  point. For a position $(R, z)$, we thus define
  $q_\text{V}(R,z) = z_{\Phi} / R_{\Phi}$, where
  $\Phi(R=0, z_{\Phi}) \equiv \Phi(R_{\Phi}, z=0) \equiv
  \Phi(R,z)$. On such equipotential, it must hold that
  $\tilde{r}(R=0, z_{\Phi}) = \tilde{r}(R_{\Phi}, z=0)$, and since
  $\tilde{r}$ only depends on $c/a$, $q_\text{V}(R,z)$ is independent
  of our mass parameter and scale radius\footnote{More precisely,
    $\tilde{r}$ depends on $r_a$ and $r_E(c/a)$, but we have chosen to
    fix the value of $r_a$ (and make it independent of $R_s$).}. We
  take values for $z_{\Phi}$ from $0.5$ to $2.0$~kpc in steps of
  $0.05$~kpc along the minor axis and compute the corresponding
  $R_{\Phi}$-values (i.e. the radii where the equipotential contours
  that belong to $z_{\Phi}$ cross the major axis). For a given $c/a$
  we then compute the mean of the absolute differences between the
  Evans mock galaxy potential flattening ($q=0.8$) and the
  Vogelsberger potential flattening along the defined range for
  $z_{\Phi}$, i.e. we compute:
  $\text{mean}(|q-q_\text{V}(R=0,z_{\Phi})|)$. We find that for
  $c/a \simeq 0.776$ this average difference is smallest (see left
  panel of Fig. \ref{fig:findingvogelsbergertruth}). For our range of
  $z_{\Phi}$ and $c/a = 0.776$, the Vogelsberger potential flattening
  increases almost linearly with $z_{\Phi}$, though the gradient is
  small ($0.018$~kpc$^{-1}$).

Given this value for the flattening, we proceed to obtain the best equivalent values for the mass and scale radius of the mock galaxy now described by the Vogelsberger profile. We do this by comparing 
$|RF_R| \equiv \sqrt{R \left| \frac{\partial}{\partial R} \Phi(R,z=0) \right|}$ along the major axis and $|zF_z| \equiv \sqrt{\left| \, z \, \frac{\partial}{\partial z} \Phi(R=0,z) \right|}$ along the minor axis with respect to their values for the mock Galaxy. We investigate their trends for $R$- and $z$-values identical to those used for $z_{\Phi}$ previously. 

We vary the scale radius %from $0.1$ to $8.0$~kpc
and the mass parameter
$\log_{10}(M_\text{1kpc}[M_{\odot}])$ %from $7.3$ to $8.0$.
and compute the mean of the absolute differences with respect to the
mock galaxy obtained along the major and minor axis for
$c/a = 0.776$. We denote this by 
$\langle \bigtriangleup v \rangle := \text{mean}[0.5 \{ \text{abs}(\Delta|RF_R|) + \text{abs}(\Delta|zF_z|) \} ]$. From the right panel of
Fig. \ref{fig:findingvogelsbergertruth} we infer that
$\langle \bigtriangleup v \rangle$ is minimum
for mass parameter
$\log_{10}(M_\text{1kpc}[M_{\odot}]) \simeq 7.69$ and scale radius
$R_s = 4.9$~kpc (green circle), although any value with
$R_s \geq 2$~kpc works well, as $\langle \bigtriangleup v \rangle$
does not vary strongly. To be able to compare these findings to the
results from our Schwarzschild models (see
Sect. \ref{subsubsec:fittingvogelsbergermodels}), we estimate the
error on these `true' parameters by considering those locations where
$\langle \bigtriangleup v \rangle$ changes by a factor 2 with respect to its
minimum value (green contour). The mass parameter is then within the
range [$7.63$, $7.73$], the scale radius larger than $2.4$~kpc. For
the smaller scale radii ($R_s < 2$~kpc) slightly higher values for the
characteristic mass parameter would be preferred, but
$\langle \bigtriangleup v \rangle$ is also larger in such cases. Note
that the NFW mass value that we just estimated corresponds well to the
mass enclosed within $1$~kpc of a spherical Evans model with
\mbox{$R_c = 1$~kpc} and \mbox{$v_0 = 20$~km/s} (as assumed in
Sect. \ref{sec:amockgalaxy}), since then
$\log_{10}(M_\text{1kpc,Evans}[M_{\odot}]) \simeq 7.67$.

Although we will not constrain the flattening of the system, we can investigate how the expected best-fit parameters change if different values for $c/a$ are taken. Setting $c/a=0.70$ results in $\langle \bigtriangleup v \rangle = 0.32$~km/s for its minimum at $\log_{10}(M_\text{1kpc}[M_{\odot}]) \simeq 7.69$ and $R_s = 4.4$~kpc, and setting $c/a=0.85$ results in $\langle \bigtriangleup v \rangle = 0.37$~km/s for $\log_{10}(M_\text{1kpc}[M_{\odot}]) \simeq 7.69$ and $R_s = 5.1$~kpc. The expected best-fit mass parameter is thus not affected by the choice of $c/a$.  The expected best-fit scale radius only increases slightly for larger values for the flattening parameter (i.e. rounder shapes), though the effect\footnote{Even for a spherical potential, i.e. $c/a=1.0$, we find the minimum $\langle \bigtriangleup v \rangle = 0.62$~km/s to be located at $\log_{10}(M_\text{1kpc}[M_{\odot}]) \simeq 7.68$ and $R_s = 6.1$~kpc.} is rather small. In addition, the grey contours, which are drawn at fixed $\langle \bigtriangleup v \rangle$, span very similar regions for different values for $c/a$.

In Fig. \ref{fig:comparingvogelsbergertruth} we compare the Vogelsberger equivalent potential to the true Evans potential of our galaxy. In the left panel we show the gradients of the potentials along the major and minor axis. Note that the Evans model seems to have 
lower $|RF_R|$ and $|zF_z|$ for $R\lesssim1$~kpc and
$z\lesssim0.75$~kpc, respectively, than the NFW `equivalent' model. In
the panel on the right we confirm that the potential flattening is
matched quite well by showing isopotential contours. Both panels reveal 
that only in the centre ($<0.7$~kpc) and at the
distances larger than $3$ kpc, the gradients of both potentials start
to deviate from each other. 

In summary, the equivalent Vogelsberger system can be described by $\log_{10}(M_\text{1kpc}[M_{\odot}]) \simeq 7.69^{+0.04}_{-0.06}$ and by $R_s \gtrsim 2.4$~kpc (with its most likely value at $R_s=4.9$~kpc) for $c/a=0.776$.

%__________________________________________________________________
%__________________________________________________________________
%__________________________________________________________________

\subsubsection{Fitting Vogelsberger models with the Schwarzschild method: exploring different sample sizes}
\label{subsubsec:fittingvogelsbergermodels}
Since we could not constrain the flattening parameter when the potential functional form was known (see Sect \ref{subsec:recoveringthemockgalaxyparameters}), we can not expect to constrain the flattening if we examine a different functional form. We set $c/a=0.80$, equal the observed flattening in the light, and subsequently find the inferences on the mass and scale radius. 
%We consider 6 different values for the flattening parameter $c/a$ ranging from $0.70$ to $0.95$ with steps of $0.05$. 
%For each value 
We initially make a grid in ($\log_{10}(M_\text{1kpc})$, $R_s$)-space,
where $R_s$ ranges from $1$ to $8$~kpc with steps of
$\Delta R_s = 1$~kpc, while for the characteristic mass
we take steps of $0.05$ for values from
$\log_{10}(M_\text{1kpc}[M_{\odot}])=7.55$ to
$\log_{10}(M_\text{1kpc}[M_{\odot}])=7.85$, i.e. just spanning a factor of
2 in mass. Later, we also decided to sample
$\log_{10}(M_\text{1kpc}[M_{\odot}])=[7.68,7.72]$ for
$R_s \in [1.5,7.5]$~kpc with a similar $\Delta R_s$ step.

\begin{figure}[t!] \centering
\includegraphics[width=0.5\textwidth]{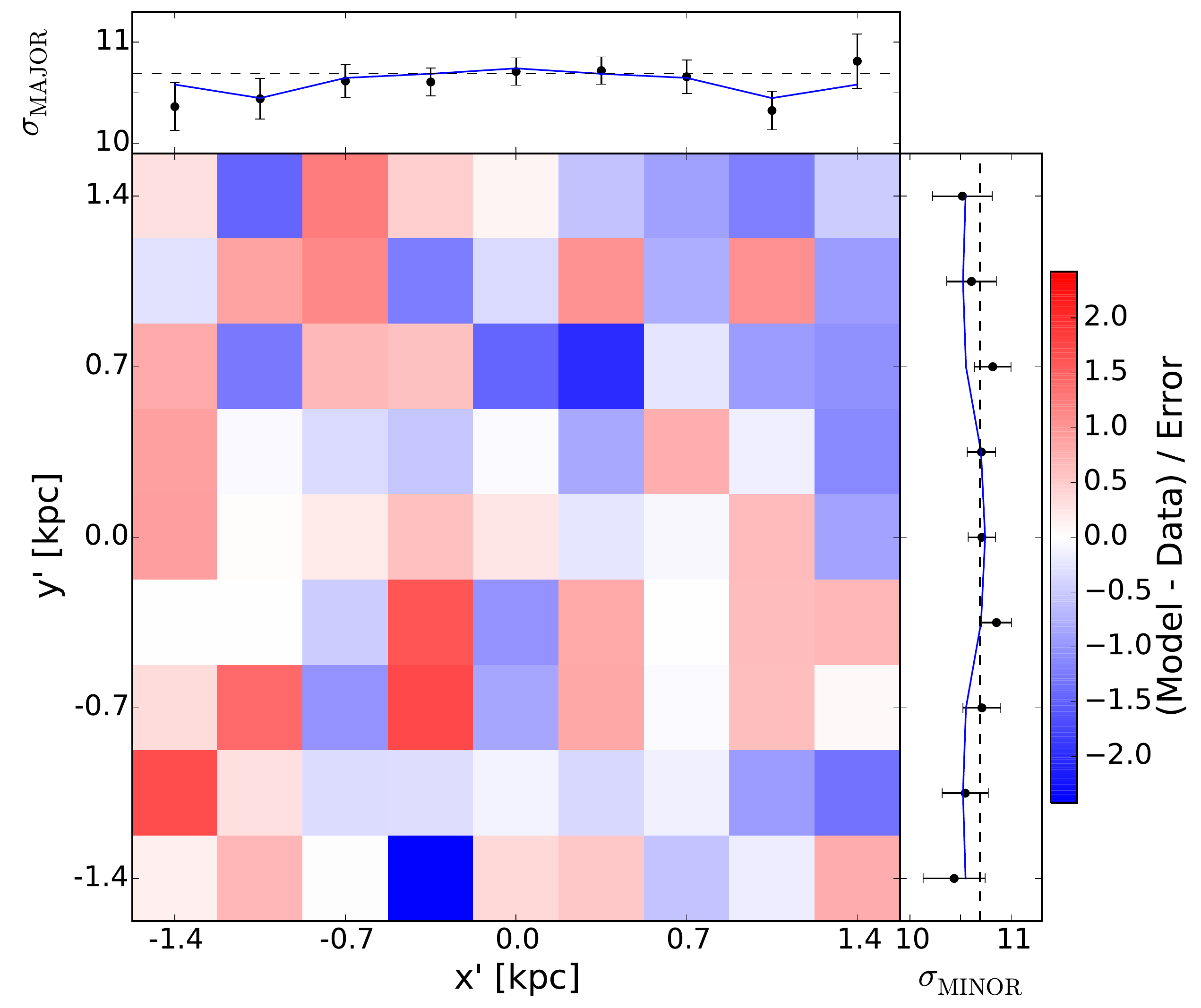}
\caption{Difference between the best fit Vogelsberger model
  (M772Rs250, blue line in the subpanels) and the observed velocity dispersion
  when applying the Schwarzschild method in 9x9 kinematic-bins to our
  mock dataset consisting of $10^5$ stars in the FOV (see
  Fig.~\ref{fig:testevans_2apertures_100000k_moments_fov_theon_results1_q80v20_chisigma}
  for a comparison).}
\label{fig:vogelsberger_chisigma}  \end{figure}

To be more efficient we decrease the number of orbits compared to
Sect. \ref{subsec:recoveringthemockgalaxyparameters} and set
$N_\text{ener}=24$, $N_{I_2}=24$ and $N_{I_3}=8$, such that a total of
$24 \times 24 \times 8 \times 5^3 = 576000$ suborbits are integrated
and $2 \times 24 \times 24 \times 8 = 9216$ orbital weights are
determined. We have found this gives good results in terms of recovery
of the light profile and kinematics.  In addition we also add
regularisation terms to the fit in this more realistic experiment: by
applying regularisation we set additional constraints such that the
orbital weights are more smoothly distributed, i.e. in a more physical
way (as the weights relate to the distribution function, which itself
is expected to be smooth). More details on the concept of
regularisation and its effects can be found in Appendix
\ref{App:AppendixB}.

We present the results following the same structure of Sect. \ref{subsec:recoveringthemockgalaxyparameters} 
and name the Vogelsberger models by MxxxRsyyy, where xxx $= 100 \log_{10}(M_{\text{1kpc}}[M_{\odot}])$ and yyy = $100 R_{s}$ (in kpc). We discuss how well we can recover the characteristic parameters of the Vogelsberger potential for mock datasets containing $10^5$, $10^4$ and $2000$ stars.

We start with the case of $10^5$ stars for which we use 9x9 kinematic-bins and no folding. For this case, we find that model M772Rs250 provides the best fit (\mbox{$\chi^2_\text{tot} = 275.1$}). Fig.~\ref{fig:vogelsberger_chisigma} shows that this model reproduces well the mock velocity dispersions in all kinematic-bins (since 
\mbox{$\chi^2_\text{kin} = 220.9$}, which results in $0.68$ per kinematic constraint). The fit is of comparable quality to the best-fit Evans model (for the same case) although the light is recovered slightly less well, which may be driven by the smaller number of orbits being used now. 

% $10^5$ stars (9x9 kinematic-bins)
\begin{figure}[t!] \centering
    \includegraphics[width=0.5\textwidth]{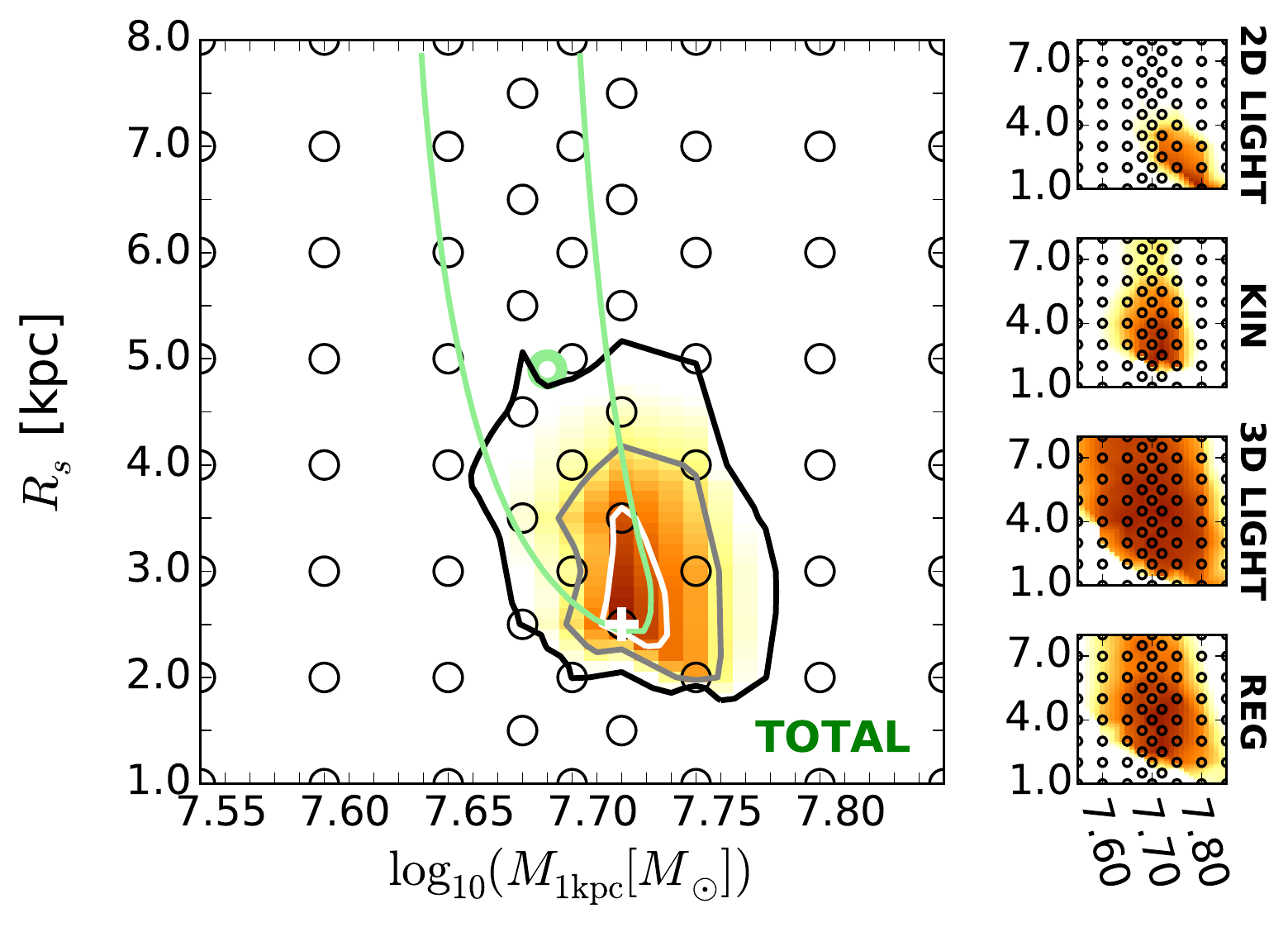}%
    \caption{Confidence intervals for the axisymmetric Vogelsberger
      model in ($\log_{10}(M_\text{1kpc})$, $R_s$) (after fixing
      $c/a = 0.8$) for the dataset with $10^5$ stars and 9x9
      kinematic-bins. The $\Delta \chi^2=[2.3, 6.18, 11.8]$-contours
      are in white, grey and black respectively. The best-fit model is
      indicated by the white cross, while the expectations are given
      by the green contour (identical to that shown in
      Fig. \ref{fig:findingvogelsbergertruth}).  The mass parameter is
      well constrained and models with $R_s \leq 2.0$~kpc are
      strongly disfavoured, consistent with our expectations. The
      small panels on the right show the $\Delta \chi^2$-landscapes
      when only considering $\chi^2_{\text{2D light}}$ (top),
      $\chi^2_{\text{kin}}$ (top-middle), $\chi^2_{\text{3D light}}$
      (bottom-middle), or $\chi^2_{\text{reg}}$ (bottom).
    } \label{fig:vogelsberger_results1_c_all_probcontours} \end{figure}

  Fig.~\ref{fig:vogelsberger_results1_c_all_probcontours} shows the
  resulting $\Delta \chi^2$-distribution in
  ($\log_{10}(M_\text{1kpc})$, $R_s$)-parameter space. The scale
  radius of the Vogelsberger potential is constrained to
  \mbox{$R_s = 2.5^{+0.6}_{-0.1}$ kpc} and the mass parameter to
  $\log_{10}(M_\text{1kpc}[M_{\odot}]) = 7.72^{+0.01}_{-0.01}$.  
  The Schwarzschild model thus prefers values towards the lower end for the
  scale radius and a mass parameter that agrees well with of our
  expectations.  The panels on the right show the
  $\Delta \chi^2$-landscapes when only considering
  $\chi^2_{\text{2D light}}$ (top), $\chi^2_{\text{kin}}$
  (top-middle), $\chi^2_{\text{3D light}}$ (bottom-middle), or
  $\chi^2_{\text{reg}}$ (bottom). The total $\Delta \chi^2$-landscape
  is dominated by the kinematics and 2D light.

\begin{figure}[t!] \centering 
    \includegraphics[width=0.50\textwidth]{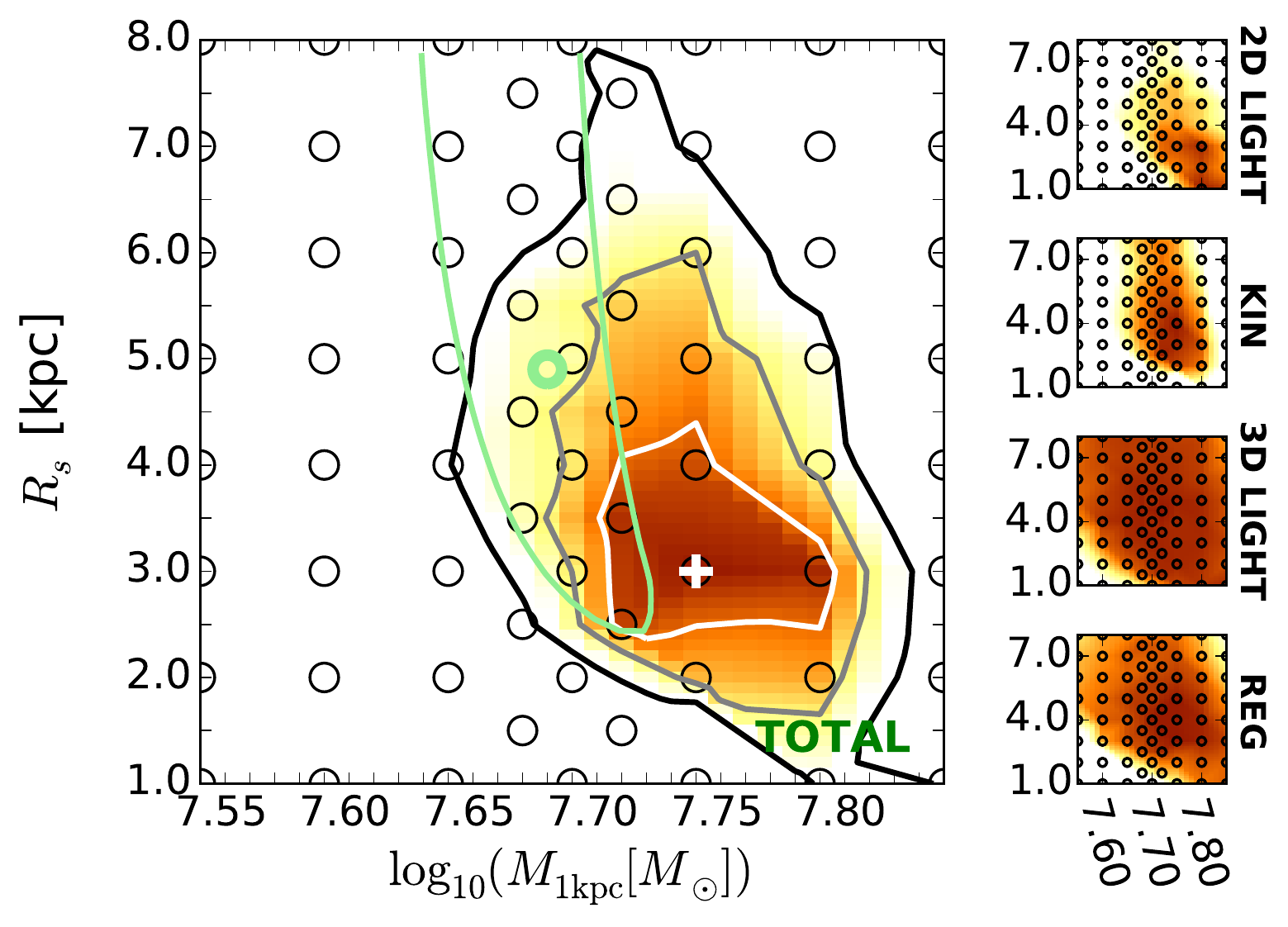}%
    \caption{Similar to
      Fig. \ref{fig:vogelsberger_results1_c_all_probcontours}, but now
      after fitting mock data consisting of $10^4$ stars and folding
      into 5x5 kinematic-bins. The decrease in sample size (by a
      factor 10) has led to a slight increase by the area spanned by
      the probability contours, although the inference on the mass
      parameter is still very good and only changed to slightly higher
      masses.}  \label{fig:vogelsberger_results2_folded9x9_c_all_probcontours}  
    \end{figure}

  Similar best-fit parameters are obtained for a smaller mock dataset
  with $10^4$ stars when folding the data into 5x5 kinematic-bins, as
  shown in
  Fig. \ref{fig:vogelsberger_results2_folded9x9_c_all_probcontours}. The
  mass and scale parameters are constrained to
  \mbox{$R_s = 3.0^{+0.7}_{-0.4}$ kpc} and
  $\log_{10}(M_\text{1kpc}[M_{\odot}]) = 7.75^{+0.05}_{-0.03}$.  For
  the best-fit model M775Rs300, \mbox{$\chi^2_\text{tot} = 78.0$} and
  \mbox{$\chi^2_\text{kin} = 33.9$}, or $0.339$ per kinematic
  constraint on average. This $\chi^2_\text{tot}$ is lower than for
  the case of $10^5$ stars, likely because we folded the data. In
  comparison to the best-fit Evans model, the quality of the fit of
  the kinematics is slightly worse but still very good.

\begin{figure}[t!] \centering 
	\includegraphics[width=0.50\textwidth]{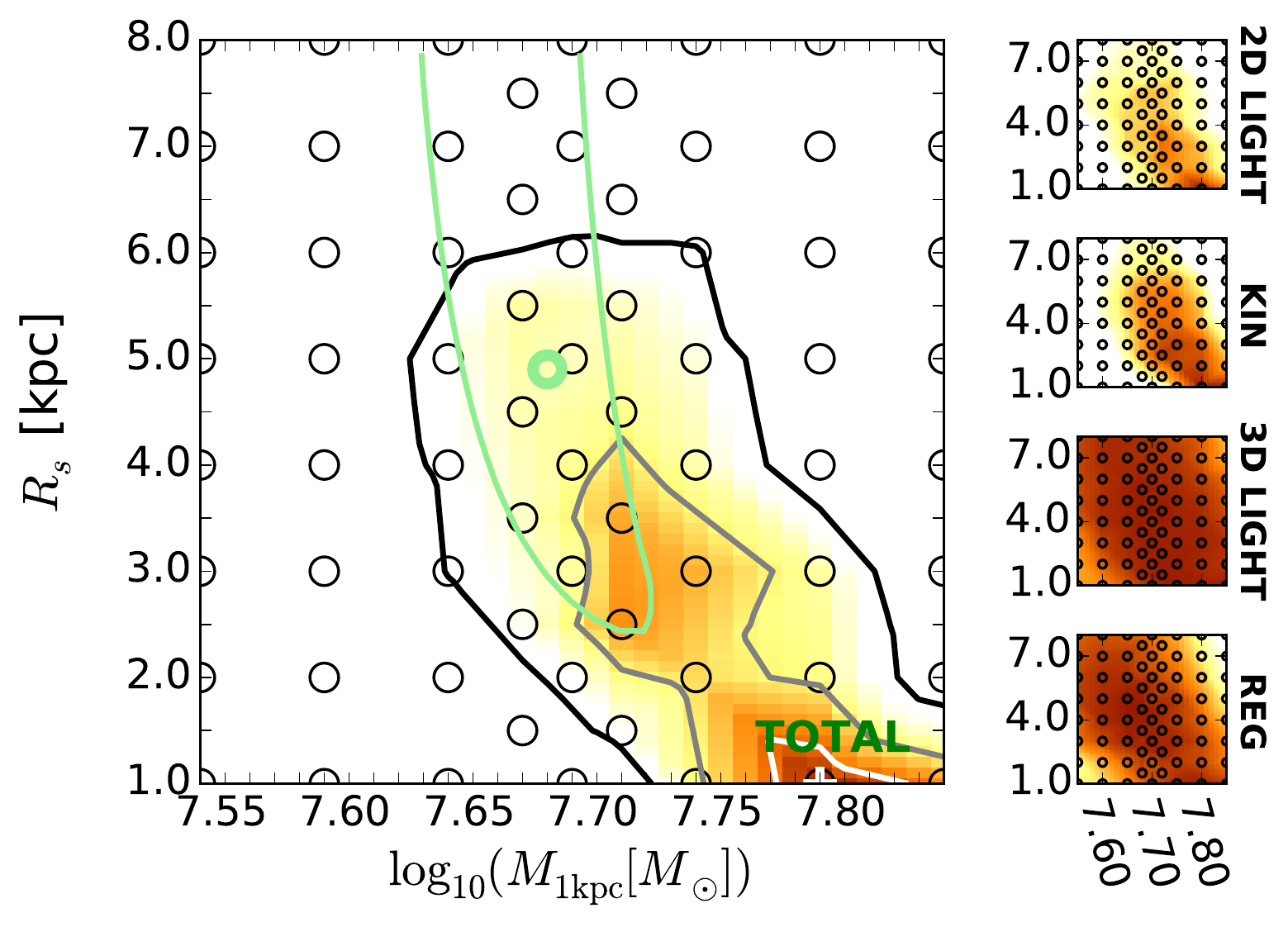}%
    \caption{
    As in Fig. \ref{fig:vogelsberger_results2_folded9x9_c_all_probcontours}, but now for a dataset with $2000$ stars. Note how the confidence contours follow the shape of the green contour (derived in Fig.~\ref{fig:findingvogelsbergertruth}).
    }  \label{fig:vogelsberger_results2000_folded9x9_c_all_probcontours} 
    \end{figure}

  When decreasing the sample size even further to $2000$ stars, we
  find that models with low values for $R_s$ and larger
  $\log_{10}(M_\text{1kpc}[M_{\odot}])$, are now preferred as shown in
  Fig.~\ref{fig:vogelsberger_results2000_folded9x9_c_all_probcontours},
  although the 95\%-confidence region still overlaps with the expected
  values for the parameters.  We obtain best-fit values of
  \mbox{$R_s = 1.0^{+0.2}_{-0.0}$ kpc} and
  $\log_{10}(M_\text{1kpc}[M_{\odot}]) = 7.80^{+0.02}_{-0.01}$. 

  It is interesting to note that the shape of the confidence contours
  obtained from the Schwarzschild method for all sample sizes, follows
  very closely the shape of the contours of $\langle \Delta v \rangle$
  depicted in Fig.~\ref{fig:comparingvogelsbergertruth}. Recall that
  the quantity $\langle \Delta v \rangle$ is a proxy for the
  difference in enclosed mass between the Evans and Vogelsberger
  model. This implies that Schwarzschild's method is actually very
  sensitive to enclosed mass, and it is identifying the set of
  Vogelsberger models that best follow the true underlying mass
  distribution. Also interesting is that the trend favouring larger
  values of the mass parameter when decreasing sample size, is present
  both for the Evans as well as for the Vogelsberger models.

% $10^4$ stars folded (9x9 into 5x5 kinematic-bins)

\begin{figure*}[t!] \centering 
	\includegraphics[width=1.0\textwidth]{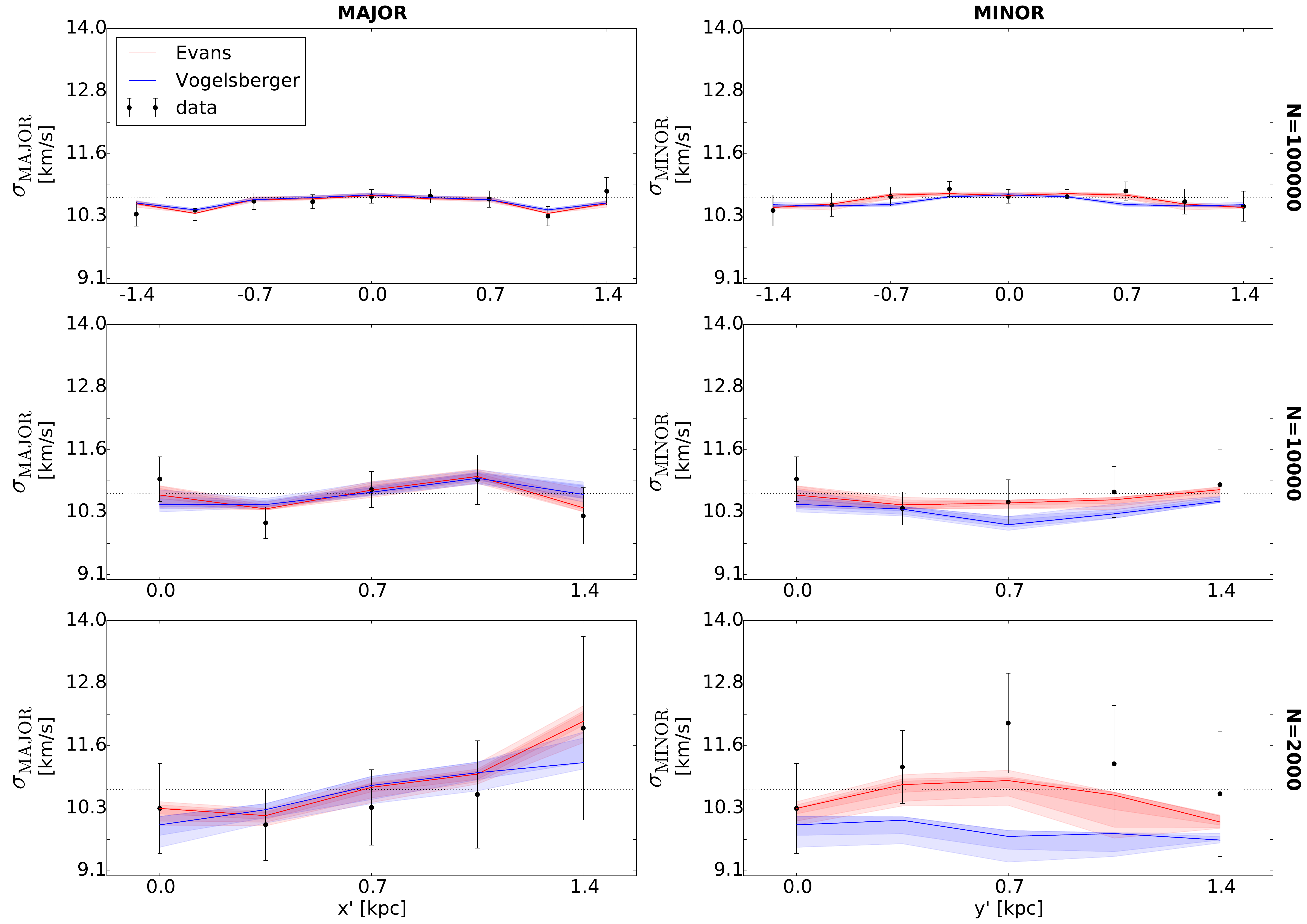}%
        \caption{Comparison of the results from the Schwarzschild
          modelling fits to the observed velocity dispersion along the
          major (left column) and minor (right column) axis. From top
          to bottom we show the best-fit Evans (red line) and
          Vogelsberger (blue line) models for datasets containing
          $N=10^5$, $N=10^4$, and $2000$ stars, respectively. The
          shaded regions denote the error bands computed as described
          in the text. Black dotted lines indicate the input
          (theoretical, Eq. \ref{eq:sigmaevans}) velocity
          dispersions.}  \label{fig:comparesigmas}
    \end{figure*}

    We compare the Evans and Vogelsberger best-fit models to the
    observed velocity dispersions in Fig. \ref{fig:comparesigmas}. The
    left and right panels compare the behaviour on the major and minor axes
    respectively, for different sample sizes: $10^5$, $10^4$ and
    $2000$ stars (in the top, middle and bottom rows
    respectively). The shaded areas enclose the minimum and maximum
    velocity dispersions for the evaluated models within the
    $\Delta \chi^2=[2.3, 6.18, 11.8]$-contours. These comparisons show
    that the Evans models fit the kinematics slightly better but that
    nearly equally good fits are provided by the Vogelsberger models (except in
    along the minor axis for the smallest dataset, bottom right panel). 

  From the analyses presented in this section we may thus conclude
  that the Schwarzschild modelling technique is sensitive to the mass
  enclosed and that it is successful in constraining well the mass parameter
  of the models, even if the functional form of the potential is not
  known.

%__________________________________________________________________
%__________________________________________________________________
%__________________________________________________________________

\section{Discussion and Conclusions}
\label{sec:discussionandconclusion}

We explored the ability of the Schwarzschild's orbit superposition
method to characterise the intrinsic properties of an axisymmetric
dSph galaxy, such as its mass, scale radius and flattening. We did
this by setting up an isothermal Sculptor-like mock galaxy that is
flattened in both the luminous and dark components. We have shown that
Schwarzschild's method applied to mock datasets with a realistic
number of stars with measured radial velocities distributed following
the luminosity profile of the system, is successful in recovering the
characteristic mass parameter of the underlying (true) logarithmic
potential, even if the potential flattening is not known. On the other
hand, we find that we can not put constraints on the flattening
parameter.

Most likely, our inability to constrain the flattening is the
consequence of our choice of the specific Evans model for our mock
galaxy. In this model with a distribution function that is ergodic,
the line-of-sight velocity profile is exactly the same everywhere and depends
on the mass parameter only. This means that the kinematics are
independent of the inclination and flattening, and the light alone
does not contain enough information to constrain the flattening
parameter. 

One might also argue that it might not be optimal for a spectroscopic
survey to sample stars according to the light profile of the
system. In fact, slightly better results were obtained when the
dataset with radial velocities provided an equal number of stars to
each kinematic-bin. All these factors, in combination with the fact
that for our specific Evans model just~$\sim\!\!30$\% of the system's
light is within our FOV, are likely playing a role. It might be
possible however that better results could be obtained with a more
realistic and general distribution function (i.e. non-ergodic),
applied to a galaxy for which the kinematic tracers cover well the
full system and sample more the outskirts.

Since in reality the potential functional form is not known, we also
explored the case in which we assume an axisymmetric NFW model. We
first determined the values of the characteristic parameters of the
NFW model that mimic the mock galaxy best by comparing some
basic properties (potential flattening and gradients in the
potential). We found that even in this case, i.e. the orbits that form
the building blocks of Schwarzschild's method are integrated in the
wrong potential, we can retrieve the correct characteristic mass and
scale parameters.

We have explored the dependencies of our results on the sizes of the
data samples used, and find that a decrease in the number of stars
with line-of-sight velocities, only slightly affects the determination
of the characteristic parameters of the model.  For the smallest
sample considered, with $2000$ stars, the inference on the mass
of the NFW ``equivalent'' model is somewhat poorer but the true value
differs by only 20\% from the best-fit and also lies within the 95\%
confidence interval.

We have checked that our results are not strongly dependent on the
choices of e.g. the number of orbits in the orbit libraries, number of
kinematic- or light-bins, and the number of velocity bins. Furthermore
we have also briefly investigated the distribution functions for the
the best-fit models, and found that, particularly when regularisation
is included, they are quite similar to the distribution function of
the mock dwarf spheroidal galaxy.

In conclusion, it is promising that the mass of our flattened system
can be recovered so well even if the flattening parameter is
unknown. This is also aligned with the results of
\citet{Kowalczyketal2018_theeffectofnonsphericity}, who applied their
spherical Schwarzschild models on non-spherical objects. To some
extent, this provides us with more confidence regarding previously
reported estimates of the mass of dSph galaxies obtained assuming
spherical symmetry.

\begin{acknowledgements}
We thank R. van den Bosch for supplying us the Schwarzschild code and L. Posti and P.T de Zeeuw for many useful discussions regarding the project. A.H. acknowledges financial support from a VICI grant from the Netherlands Organisation for Scientific Research, NWO. 
\end{acknowledgements}

%-------------------------------------------------------------------

\bibliographystyle{aa}
\bibliography{bibliography}

%__________________________________________________________________
%__________________________________________________________________
%__________________________________________________________________

\begin{appendix}

\section{\\Generating a mock dataset with realistic errors} \label{App:AppendixA}
% the \\ insures the section title is centered below the phrase: AppendixA
Like in \citet{Breddelsetal2013}, we define $v_i$ as the true
line-of-sight velocity of star $i$ and $\epsilon_i$ as the (true and
unknown) measurement error on that star. Therefore $v_i + \epsilon_i$
is the observed velocity of star $i$. We note that the expectation
values for the moments of the measurement errors, which are drawn from
a Gaussian distribution with $\sigma=2$ km/s, are given by:
$E\left[ \langle \epsilon^n_i \rangle \right]= E \left[ \epsilon^n_i
\right] = 0$ for odd $n$ and
$s_n \equiv E \left[ \langle \epsilon^n_i \rangle \right] = E
\left[\epsilon^n_i \right] = (n-1)!!\sigma^n$ for even $n$.  In our
terminology, $\hat{\mu_n} = E[\langle v_i^n \rangle] = E[v_i^n]$
denotes the true $n^{\text{th}}$ moment, $\mu_n$ is its estimator and
the observed $n^{\text{th}}$ moment is
$m_n = \frac{1}{N} \sum\limits^N_{i=1} (v_i + \epsilon_i)^n$ for a
sample of $N$ stars in a given positional bin on the sky
(i.e. kinematic-bin). 

Since we want to know the true value of the moments, i.e. without
measurement errors, we will compute the estimators of the true
moments. We will also use raw moments (i.e. not taken about the mean
velocities), and in what follows, we thus refer to `moments' to denote
`raw moments'. Since we can only in practise compute the estimators of
the true moments, we replaced $\hat{\mu_n}$ by $\mu_n$ in the
right-hand side of the following equations. The first four moment
estimators are then given by:
\begin{equation}
\mu_1 = \frac{1}{N} \sum\limits^N_{i=1} (v_i + \epsilon_i) \, ,
\end{equation}
\begin{equation}
\mu_2 = \frac{1}{N} \sum\limits^N_{i=1} (v_i + \epsilon_i)^2 - s_2 \, ,
\end{equation}
\begin{equation}
\mu_3 = \frac{1}{N} \sum\limits^N_{i=1} (v_i + \epsilon_i)^3 - 3 \mu_1 s_2 \, ,
\end{equation}
and 
\begin{equation}
\mu_4 = \frac{1}{N} \sum\limits^N_{i=1} (v_i + \epsilon_i)^4 - 6 \mu_2 s_2 - 3 s^2_2 \, .
\end{equation}

To compute the error on these moments, we compute the square root of the variance of the moments; $Var(\hat{\mu_n}) \approx Var(\mu_n) \approx Var(m_n) = E[{m_n}^2] - (E[m_n])^2$:
\begin{equation}
\begin{split}
Var(\mu_1) &= \frac{\mu_2 + s_2 - \mu_1^2}{N} \\
         &= \frac{1}{N} \left\{ \frac{1}{N} \sum\limits^N_{i=1} (v_i + \epsilon_i)^2 - \left[ \frac{1}{N} \sum\limits^N_{i=1} (v_i + \epsilon_i) \right]^2 \right\} \, ,
\end{split}
\end{equation}
\begin{equation}
\begin{split}
Var(\mu_2) &= \frac{1}{N} \left[ \mu_4 - \mu_2^2 + 4 \mu_2 s_2 + 2s^2_2\right] \\
         &= \frac{1}{N} \left\{ \frac{1}{N} \sum\limits^N_{i=1} (v_i + \epsilon_i)^4 - \left[ \frac{1}{N} \sum\limits^N_{i=1} (v_i + \epsilon_i)^2 \right]^2 \right\} \, ,
%&=...in terms of moments) \\
%&= .. (in terms of observed velocities!)
\end{split}
\end{equation}
\begin{equation}
\begin{split}
Var(\mu_3) &= \frac{1}{N} \left[ \mu_6 +15\mu_4 s_2 + 45\mu_2 s^2_2 + 15s^3_2 - \mu^2_3 \right.\\
         & \textrm{\, \, \,} \left. -6 \mu_3 \mu_1 s_2 - 9\mu^2_1 s^2_2 \right] \\
         &= \frac{1}{N} \left\{ \frac{1}{N} \sum\limits^N_{i=1} (v_i + \epsilon_i)^6 - \left[ \frac{1}{N} \sum\limits^N_{i=1} (v_i + \epsilon_i)^3 \right]^2 \right\} \, ,
\end{split}
\end{equation}
and 
\begin{equation}
\begin{split}
Var(\mu_4) &= \frac{1}{N} \left[ \mu_8 + 28\mu_6 s_2 -\mu^2_4 - 12 \mu_4 \mu_2 s_2 + 204\mu_4 s^2_2  \right. \\
         & \textrm{\, \, \,} \left. - 36\mu^2_2 s^2_2 + 384\mu_2 s^3_2 + 96s^4_2\right] \\
         &= \frac{1}{N} \left\{ \frac{1}{N} \sum\limits^N_{i=1} (v_i + \epsilon_i)^8 - \left[ \frac{1}{N} \sum\limits^N_{i=1} (v_i + \epsilon_i)^4 \right]^2 \right\} \, .
\end{split}
\end{equation}
where the errors on the third and fourth moment estimators also depend on: 
\begin{equation}
\mu_6 = \frac{1}{N} \sum\limits^N_{i=1} (v_i + \epsilon_i)^6 -  15 \mu_4 s_2 -45 \mu_2 s^2_2 - 15 s^3_2 \, ,
\end{equation}
and
\begin{equation}
\mu_8 = \frac{1}{N} \sum\limits^N_{i=1} (v_i + \epsilon_i)^8 - 28 \mu_6 s_2 - 210 \mu_4 s^2_2 - 420 \mu_2 s^3_2 - 105 s^4_2 \, .
\end{equation}
Obviously the errors on the moments decrease when the number of stars in a kinematic-bin increases.\\

%__________________________________________________________________
%__________________________________________________________________
%__________________________________________________________________

\section{The effect of the sampling of line-of-sight velocities} \label{app:light} 

\begin{figure*}[ht!] \centering
	\includegraphics[width=0.45\textwidth]{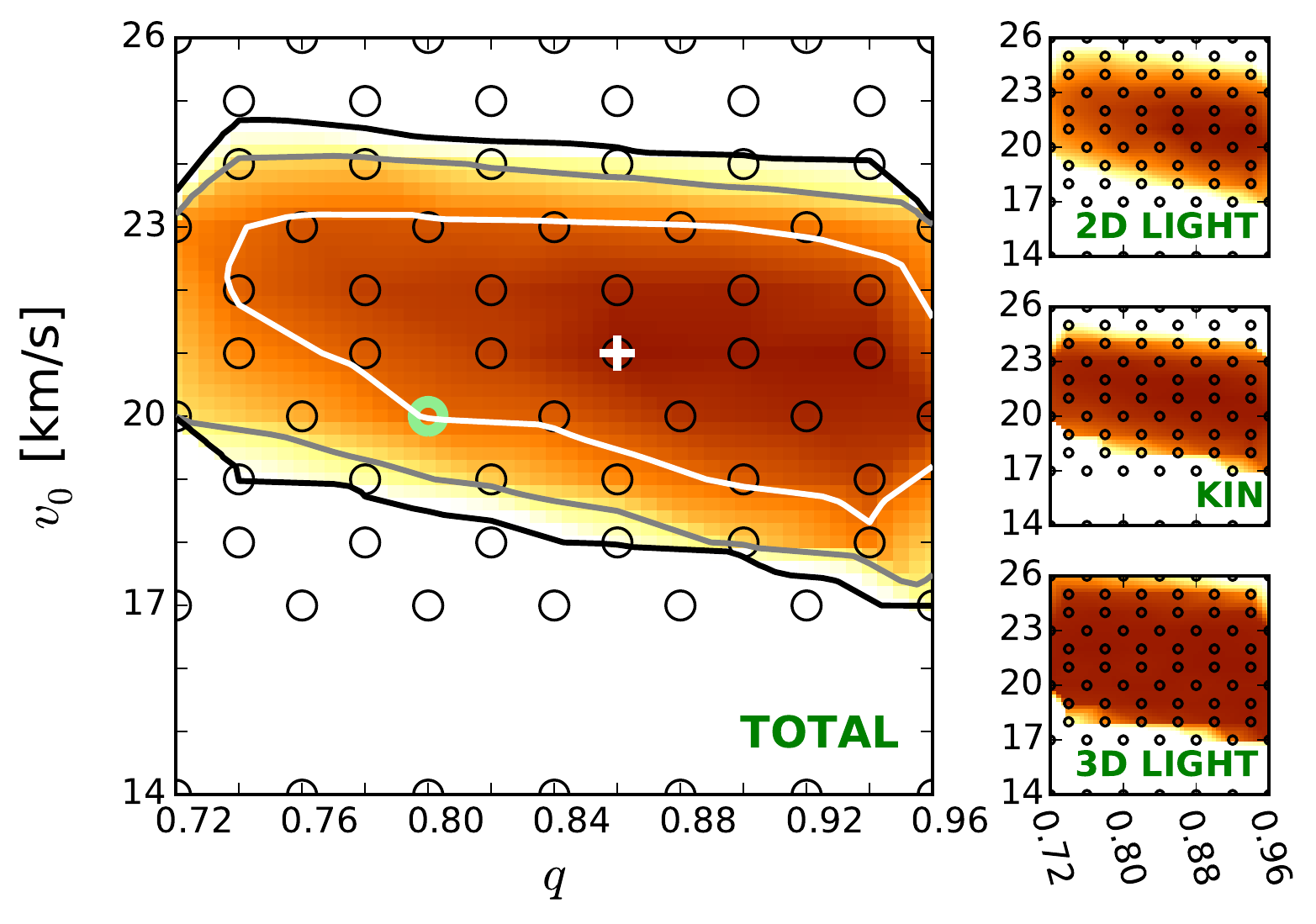}%
    \includegraphics[width=0.45\textwidth]{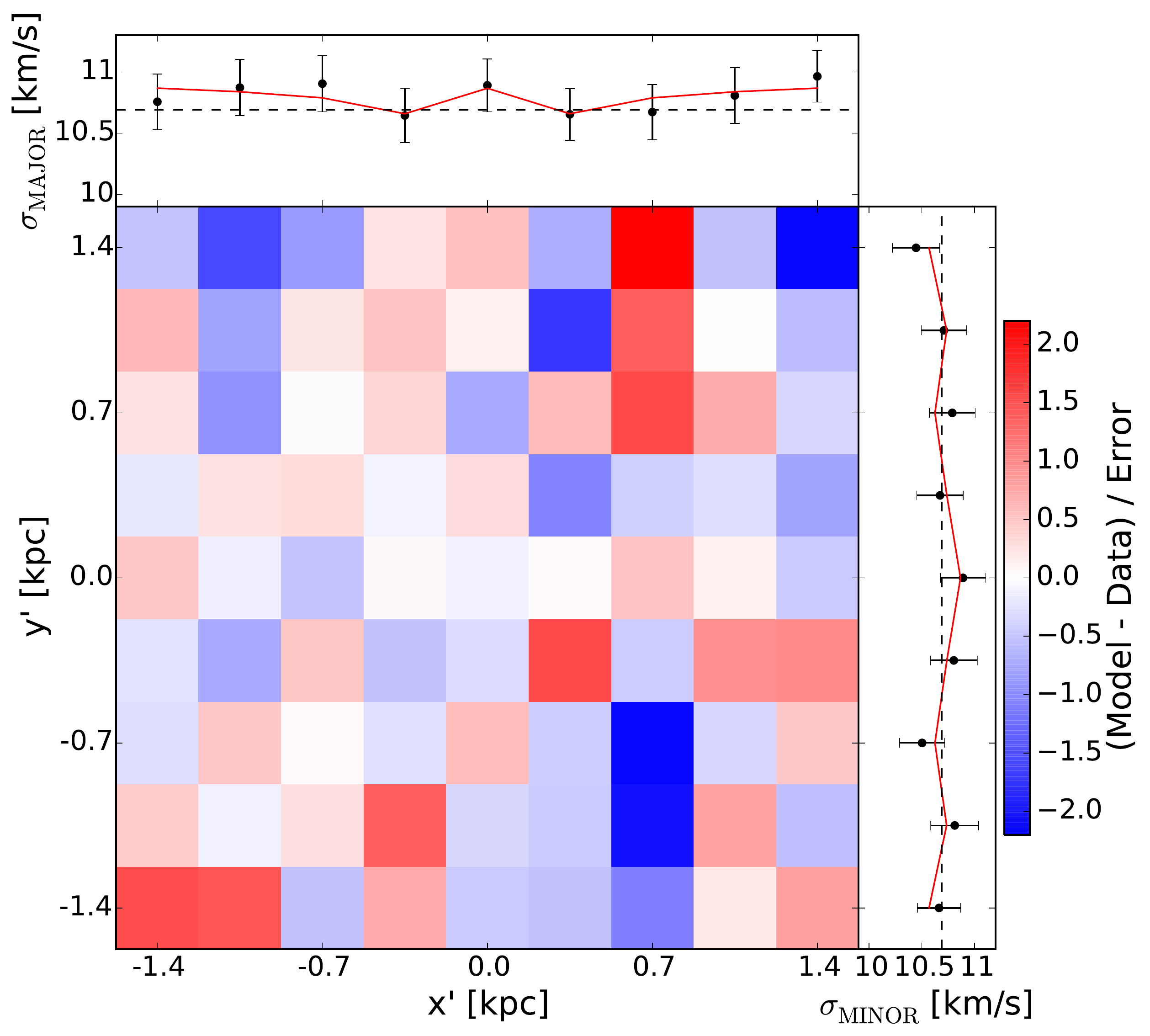}%
	\caption{Left: The confidence intervals after fitting the Evans models to a new realization of the dataset containing $10^5$ stars. This time we ensure an equal number of stars per kinematic-bin. The inference on the mass parameter remains the same. The flattening parameter remains unconstrained but has slightly shifted into the direction of the correct flattening parameter.
	Right: The velocity dispersion profile for the best-fit model q86v21.
	} \label{fig:resultsENSPB} \end{figure*}%

In the main paper we have drawn samples of line-of-sight velocities
that follow the light distribution of the mock galaxy. Here we show
the results of applying the Schwarzschild modelling technique to a
dataset consisting of $10^5$ stars, but this time distributed such
that each kinematic-bin has an equal number of stars. 

Fig.~\ref{fig:resultsENSPB} presents the inference on the mass and
flattening parameter and should be compared to
Fig.~\ref{fig:evans_results1_probcontours} and
\ref{fig:testevans_2apertures_100000k_moments_fov_theon_results1_q80v20_chisigma}. As
can be observed, we have very similar inferences with the best-fit
flattening parameter slightly moved into the direction of the
input/true flattening value. Nonetheless, this remains fairly
unconstrained.

%
%__________________________________________________________________
%__________________________________________________________________
%__________________________________________________________________

\section{Regularisation} \label{App:AppendixB}
% the \\ insures the section title is centered below the phrase: AppendixB

\begin{figure*}[ht!] \centering
    \includegraphics[width=0.8\textwidth]{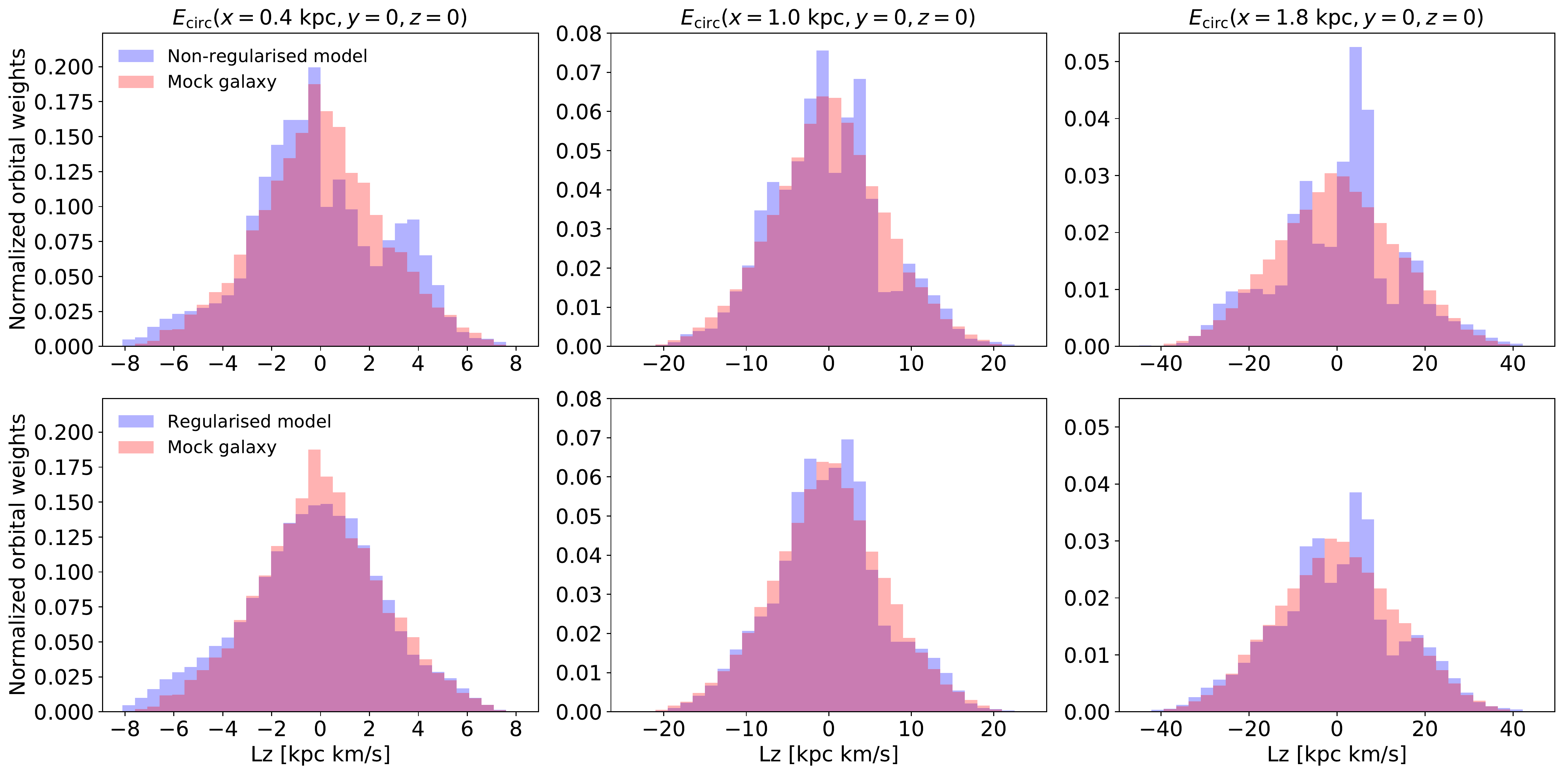}%
    \caption{Orbital distributions of angular momentum around the
      symmetry axis, i.e. $L_z$ for fixed energy slices corresponding
      to circular orbits at $x=0.4$ (left), $x=1.0$ (middle), and
      $x=1.8$~kpc (right). Note the different axes ranges for the
      panels of each column. In the top row we show the distributions
      for the true Evans model q80v20 (blue) obtained for a dataset of
      $10^5$ stars (see
      Sect. \ref{subsec:recoveringthemockgalaxyparameters}) and for a
      realization of the mock galaxy (red, here containing
      $4\times 10^5$ stars in total). The effect of adding
      regularisation to the fit is shown in the bottom panels. Adding
      regularisation makes the recovered distribution smoother and more similar
      to the true distribution.
    } \label{fig:df_q80v20_addingregularisation} \end{figure*}%
    
The solution of our minimisation problem may result in a distribution of orbital weights that is rapidly varying or shows sharp discontinuities. Such a distribution would not be physical. Therefore we make the distribution of the orbital weights smoother by adding extra terms to the $\chi^2$-fitting algorithm such that a new quantity $\widetilde{\chi^2_{\text{tot}}}$ is minimised:
\begin{equation}
\label{eq:chi2totplusreg}
\widetilde{\chi^2_{\text{tot}}} = \chi^2_{\text{tot}} + \chi^2_{\text{reg}} \, .
\end{equation}
This procedure is called regularisation. The regularisation strength is chosen such that the orbital weights are forced to change smoothly from one neighbouring orbit to the next, while finding similar values for the best-fit characteristic parameters. In addition, the confidence contours should not be significantly shaped by the $\chi^2_{\text{reg}}$-term. We refer the reader to \citet{vandenBoschetal2008} for more information about the exact implementation, in particular to Eqs. 28 and 29 of that paper. These equations require the 3-dimensional stellar density profile. For this work we assumed to know $\rho_{\text{lum}}$ (see Eq. \ref{eq:rholum}). In reality one needs the inclination angle to transform the observed surface brightness profile into the stellar density profile. 

In the bottom panels of Fig. \ref{fig:df_q80v20_addingregularisation},
we show the effect of adding regularisation for the Evans q80v20 model
(i.e. this is the true model) on the distribution of angular momentum
around the symmetry axis ($L_z$). The distributions can be compared to
those of the q80v20 model without regularisation (top rows). We here
show the example with $10^5$ stars with line-of-sight velocities. The
modelled distribution functions (blue) are generally smoother when
regularisation is used. As a reference we also include the
distributions for a realization of the mock galaxy (red). The model
reproduces the mock distribution reasonably well, though some differences
exist. The fact that only $\sim30\%$ of the
total number stars of the mock galaxy end up in our FOV might play a
role here, in addition to the fact that we have discretized the data
(by using kinematic-bins, and by modelling only the first four
velocity moments).

\end{appendix}

\end{document}